\documentclass[a4paper,fleqn,usenatbib]{mnras}

\usepackage{newtxtext,newtxmath}

\usepackage[T1]{fontenc}
\usepackage{ae,aecompl}

\usepackage{graphicx}	
\usepackage{amsmath}	
\usepackage{amssymb}	

\title[Mean density inversions for giant stars]{Mean density inversions for red giants and red clump stars}

\author[G. Buldgen et al.]{
Ga\"el Buldgen$^{1}$\thanks{E-mail: G.Buldgen@bham.ac.uk},
B. Rendle$^{1}$,
T. Sonoi$^{2,3}$,
G.R. Davies$^{1}$,
A. Miglio$^{1}$,
S.J.A.J. Salmon$^{4}$,
\and
D. R. Reese$^{3}$,
D. Bossini$^{5}$,
P. Eggenberger$^{6}$
A. Noels$^{3}$, and
R. Scuflaire$^{3}$ 
\\
$^{1}$School of Physics and Astronomy, University of Birmingham, Edgbaston, Birmingham B15 2TT, UK. \\
$^{2}$Astronomical Institute, Tohoku University, 6-3 Aramaki Aza-Aoba, Aoba-ku Sendai, 980-8578, Japan \\
$^{3}$LESIA, Observatoire de Paris, Universit\'e PSL, CNRS, Sorbonne Universit\'e, Univ. Paris Diderot, Sorbonne Paris Cit\'e,\\ 5 place Jules Janssen, 92195 Meudon, France. \\
$^{4}$STAR Institute, Universit\'e de Li\`ege, All\'ee du Six Ao\^ut 19C, B-4000 Li\`ege, Belgium.\\
$^{5}$Observatory of Padua-INAF, Vicolo dell'Osservatorio 5, I-35122 Padova, Italy. \\
$^{6}$Observatoire de Gen\`eve, Universit\'e de Gen\`eve, 51 Ch. Des Maillettes, CH$-$1290 Sauverny, Suisse
}

\date{Accepted XXX. Received YYY; in original form ZZZ}

\pubyear{2018}

\begin{document}
\label{firstpage}
\pagerange{\pageref{firstpage}--\pageref{lastpage}}
\maketitle

\begin{abstract}
Since the CoRoT and Kepler missions, the availability of high quality seismic spectra for red giants has made them the standard clocks and rulers for Galactic Archeology. With the expected excellent data from the TESS and PLATO missions, red giants will again play a key role in Galactic studies and stellar physics, thanks to the precise masses and radii determined by asteroseismology. The determination of these quantities is often based on so-called scaling laws, which have been used extensively for main-sequence stars. We show how the SOLA inversion technique can provide robust determinations of the mean density of red giants within 1 per cent of the real value, using only radial oscillations. Combined with radii determinations from Gaia of around 2 per cent precision, this approach provides robust, less model-dependent masses with an error lower than 10 per cent. It will improve age determinations, helping to accurately dissect the Galactic structure and history. We present results on artificial data of standard models, models including an extended atmosphere from averaged 3D simulations and non-adiabatic frequency calculations to test surface effects, and on eclipsing binaries. We show that the inversions provide very robust mean density estimates, using at best seismic information. However, we also show that a distinction between red-giant branch and red-clump stars is required to determine a reliable estimate of the mean density. The stability of the inversion enables an implementation in automated pipelines, making it suitable for large samples of stars.
\end{abstract}




\section{Introduction} \label{sec:intro}
 
Red giants play a key role in stellar physics. Since the detection of mixed modes in their oscillation spectra \citep{DeRidder} thanks to the CoRoT \citep{Baglin} and \textit{Kepler} \citep{Borucki} missions, they are at the origin of multiple questions on the reliability of our depiction of stellar structure and evolution \citep{Mosser,Deheuvels}. The availability of thousands of high quality seismic spectra for these stars led to their use as the standard clocks and rulers \citep{MiglioClocks} for Galactic Archeology \citep{MiglioPopulations,Anders2017RG}. Today, they are used as tracers of the structure and chemical evolution of our Galaxy \citep{AndersRG}. New accurate data for these stars are also expected  to be delivered by the TESS and PLATO \citep{Rauer} missions, which will play a key role in Galactic studies \citep{Miglio2017}. These successes originate from the ability of asteroseismology to provide precise masses and radii for a large number of stars. The seismic determination of these quantities is often based on so-called scaling laws, which have also been used extensively for main-sequence stars.

However, while the precision of these determinations is excellent, due to the high precision of the space-based photometry data, their accuracy is far from perfect. Multiple studies \citep{Gaulme2016,Brogaard2016,Rodrigues2017, Viani2017, Brogaard2018} have shown that they could lead to inaccurate results. From a physical point of view, their limited accuracy is not surprising as they do not fully exploit the information contained in the seismic spectra. Therefore, providing a more robust way of determining the mean density of the observed targets using seismology is required so that, using constraints from the second Gaia data release, more accurate masses can be determined for thousands of stars. These accurate masses will help with dissecting the structure of the Galaxy, thus providing new insights on its evolution and formation history. In addition to this potential, the determination of accurate fundamental parameters of red giants in stellar clusters is also crucial to constrain the mass loss rate on the red giant branch, a still uncertain key phenomenon of stellar evolution \citep{MiglioClusters, HandbergRG}.

In this study, we will show how the adaptation of the SOLA inversion technique for the mean density developed by \citet{Reese} used on the radial oscillations of red giant stars can provide more robust determinations of the mean density than values obtained from the fitting of the average large frequency separation or the usual scaling laws. In section \ref{sec:inversion}, we briefly recall the principles of the inversion techniques. In section \ref{sec:numex}, we test the inversion in various numerical exercises, using artificial targets on the red giant branch (hereafter denoted RGB), in the red clump, and an RGB target including an averaged $3$D atmosphere model for which the frequencies are computed using adiabatic and non-adiabatic oscillation codes to test various surface effects correction. In section \ref{sec:ObsBin}, we apply our method to a subsample of eclipsing binaries studied by \citet{Gaulme2016} and \citet{Brogaard2018}. This is then followed by a conclusion.
 
\section{The inversion technique}\label{sec:inversion}

The inversion procedure used to obtain the mean density is that of \citet{Reese}. We only briefly recall a few specific aspects of the method for the sake of clarity. 

The goal of the approach is to determine through the SOLA inversion technique \citep{Pijpers} an estimate of the mean density of a given observed star using the linear integral structural relations between individual relative frequency differences and corrections of thermodynamic quantities such as density, $\rho$, the squared adiabatic sound speed, $c^{2}$ or the adiabatic exponent, $\Gamma_{1}=\frac{\partial \ln P}{\partial \ln \rho}\vert_{S}$, with $P$ the pressure and $S$ the entropy \citep{Dziembowski, Gough}. This can be done by using the linear perturbation of the mean density with the integral formula of the stellar mass
\begin{align}
\frac{\delta \bar{\rho}}{\bar{\rho}}=\frac{3}{4\pi R^{3}\bar{\rho}}\int_{0}^{R}4\pi r^{2}\delta \rho dr, \label{eq:EqBasisMeanDens}
\end{align}
to define the target function of the SOLA inversion. Using a little algebra in Eq. \ref{eq:EqBasisMeanDens} and by non-dimensionalising the integral, one has as
\begin{align}
\mathcal{T}_{\bar{\rho}}=4\pi x^{2} \frac{\rho}{\rho_{R}},
\end{align}
with the radial position of a layer of stellar material normalized by the photospheric stellar radius $x=r/R$, $\rho$ the density of stellar material and $\rho_{R}=M/R^{3}$, with $M$ the stellar mass and $R$ the photospheric stellar radius. Using this target, the cost function of the SOLA method becomes
\begin{align}
\mathcal{J}_{\bar{\rho}}(c_{i})=&\int_{0}^{1}\left[K_{\mathrm{Avg}} - \mathcal{T}_{\bar{\rho}}\right]^{2}dx + \beta \int_{0}^{1}\left( K_{\mathrm{Cross}}\right)^{2}dx \nonumber \\ &+ \lambda \left[ 2 -\sum_{i}c_{i} \right] + \tan \theta \frac{\sum_{i}\left(c_{i}\sigma_{i}\right)^{2}}{<\sigma^{2}>}, \label{eq:CostSOLA}
\end{align}
where we have introduced the averaging and cross-term kernels, defined as follows
\begin{align}
K_{\mathrm{Avg}}=\sum_{i}c_{i}K^{i}_{\rho,\Gamma_{1}}, \\
K_{\mathrm{Cross}}=\sum_{i}c_{i}K^{i}_{\Gamma_{1},\rho},
\end{align}
and the parameters $\beta$ and $\theta$ which define the trade-off problem between the fit of the target, the contribution of the cross term and the amplification of observational error bars of the individual frequencies, denoted $\sigma_{i}$. The $K^{i}_{\rho,\Gamma_{1}}$ and $K^{i}_{\Gamma_{1},\rho}$ are the so-called structural kernel functions, derived from the variational analysis of the pulsation equations and $<\sigma^{2}>=\frac{1}{N}\sum_{i=1}^{N}\sigma^{2}_{i}$ with $N$ the number of observed frequencies. In Eq. \ref{eq:CostSOLA}, we have also introduced the inversion coefficients $c_{i}$ and $\lambda$, a Lagrange multiplier. The third term is based on homologous reasoning described in \citet{Reese} which also leads to a non-linear generalization of the method where the inverted mean density, $\bar{\rho}_{\mathrm{Inv}}$ , is determined using the formula
\begin{align}
\bar{\rho}_{\mathrm{Inv}}=\left(1+\frac{1}{2}\sum_{i}c_{i}\frac{\delta \nu_{i}}{\nu_{i}}\right)^{2}\bar{\rho}_{\mathrm{Ref}},
\end{align}
with $\bar{\rho}_{\mathrm{Ref}}$ the mean density of the reference model of the inversion and $\frac{\delta \nu_{i}}{\nu_{i}}$ the relative differences between the observed and theoretical frequencies defined as $\frac{\nu_{\mathrm{Obs}}-\nu_{\mathrm{Ref}}}{\nu_{\mathrm{Ref}}}$. In the following sections, we will always use this non-linear generalization. Using this approach, the errors on the inverted mean density are given by
\begin{align}
\sigma_{\bar{\rho}_{\mathrm{Inv}}}=\bar{\rho}_{\mathrm{Ref}}\left(1+\frac{1}{2}\sum_{i}c_{i}\frac{\delta \nu_{i}}{\nu_{i}} \right)\sqrt{\sum_{i}c^{2}_{i}\sigma^{2}_{i}}.
\end{align}

A few additional comments can be made on Eq. \ref{eq:CostSOLA}. We have intentionally dropped the classical surface term commonly used in helioseismology \citep[see for example][]{Rabello}. In Section \ref{sec:SurfEff}, we will comment on this choice and discuss in more details the optimal approach to implement surface corrections and provide examples using the formulation of \citet{Ball1} and \citet{Sonoi} for the behaviour of the surface term for patched models and frequencies including non-adiabatic effects.

We also note that we make the choice of using the $\left( \rho,\Gamma_{1} \right)$ structural pair to carry out the inversions. Other pairs, such as the $\left(\rho,c^{2}\right)$ or the $\left( \rho,Y \right)$ pair, with $Y$ the helium mass fraction, could be used, but the former shows strong contributions from the cross-term kernel and is thus inadequate, whereas the latter requires an implementation of the equation of state which leads to accurate derivations of state derivatives of $\Gamma_{1}$ and leads to the same accuracy as the $\left( \rho, \Gamma_{1} \right)$ pair. 

The reliability of the inversion procedure is usually assessed in terms of the norm of the averaging and cross-term kernels, defined as follows
\begin{align}
\vert \vert K_{\mathrm{Avg}} \vert \vert^{2}&=\int_{0}^{1}\left[ K_{\mathrm{Avg}}-\mathcal{T}_{\bar{\rho}}\right]^{2}dx, \\
\vert \vert K_{\mathrm{Cross}} \vert \vert^{2}&=\int_{0}^{1} K_{\mathrm{Cross}}^{2}dx.
\end{align}
In addition to this analysis, it can be more thoroughly assessed when the method is applied to artificial data with the help of three other quantities. The definition of these error contributions is
\begin{align}
\epsilon_{\mathrm{Avg}}=&-\int_{0}^{1}\left( \mathcal{T}_{\bar{\rho}}-K_{Avg}\right)\frac{\delta \rho}{\rho}dx, \label{eqAvgErr} \\
\epsilon_{\mathrm{Cross}}=&-\int_{0}^{1} K_{\mathrm{Cross}} \frac{\delta \Gamma_{1}}{\Gamma_{1}}dx, \label{eqCrossErr}\\
\epsilon_{\mathrm{Res}}=&\frac{\bar{\rho}_{\mathrm{Inv}}-\bar{\rho}_{\mathrm{Obs}}}{\bar{\rho}_{\mathrm{Ref}}}-\epsilon_{\mathrm{Avg}}-\epsilon_{\mathrm{Cross}}, \label{eqResErr}
\end{align}
where we have defined $\epsilon_{\mathrm{Avg}}$, the contribution from the inaccurate fit of the target by the averaging kernel, $\epsilon_{\mathrm{Cross}}$, the error contribution from the non-zero value of the cross-term kernel and $\epsilon_{\mathrm{Res}}$, the residual error, which contains all other sources of uncertainties such as non-linear contributions, surface effects, linearization of the equation of state or systematic errors in the values of the observed frequencies. We have also used the notations  $\bar{\rho}_{\mathrm{Ref}}$ and $\bar{\rho}_{\mathrm{Obs}}$ for the mean density of the reference model and the ``observed'' artificial target respectively.

\section{Numerical exercises}\label{sec:numex}

To test the reliability and robustness of the inversion technique, we defined a set of artificial targets that would be fitted using various constraints to define reference models for the inversion. A first set of targets on the RGB and their properties are summarized in Table \ref{tabTargets}. 

\begin{table*}
\caption{Properties of the target models used for the inversion.}
\label{tabTargets}
  \centering
    \resizebox{\linewidth}{!}{%
\begin{tabular}{r | c | c | c | c | c | c | c | c | c | c | c | c }
\hline
& Mass (M$_{\odot}$)& Age (Gy)& $R$ (R$_{\odot}$)& $L$ (L$_{\odot}$)& EOS& Opacities& $\alpha_{\mathrm{MLT}}$& Diffusion& $\alpha_{\mathrm{Ov}}$&$X_{0}$&$Z_{0}$ & Mixture  \\
\hline
Target 1& $1.4$ & $3.5$ & $8.97$ & $34.2$ & FreeEOS$^{1}$ & OPAL$^{5}$ & $1.8$ & Thoul$^{8}$& $0.10$ & $0.72$ & $0.0135$ & AGSS$09^{3}$\\ 
Target 2& $1.3$ & $5.84$ & $6.75$ & $17.64$ & FreeEOS & OPAL & $1.7$ & $/$ & $0.15$ & $0.71$ & $0.022$ & AGSS$09$\\ 
Target 3& $1.05$& $10.2$ & $14.52$ & $72.48$ & FreeEOS & OPLIB$^{6}$ & $2.0$ & Paquette$^{9}$& $0.00$ & $0.71$ & $0.020$ &GN$93^{4}$\\ 
Target 4& $1.25$ & $3.96$ & $29.0$ & $270.0$ & FreeEOS & OPAL & $2.0$ & $/$& $0.0$ & $0.71$ &$0.010$ &GN$93$\\ 
Target 5& $0.98$ & $13.9$ & $31.47$ & $177.9$ & CEFF$^{2}$ & OPAS$^{7}$ & $1.5$ & Paquette& $0.10$ & $0.72$ & $0.016$ &AGSS$09$ \\ 
Target 6& $1.15$ & $7.89$ & $ 47.26$ & $418.0$ & FreeEOS & OPLIB & $1.8$ & Paquette & $0.10$ & $0.73$& $0.0155$& AGSS$09$\\
Target 7& $1.37$ & $4.98$ & $15.38$ & $80.33$ & FreeEOS & OPLIB & $1.9$ & $/$ & $0.15$ & $073$ & $0.022$ &GN$93$\\
Target 8& $2.5$ & $1.08$ & $36.78$ & $366.5$ & CEFF & OPLIB & $1.7$ & $/$ & $0.15$ & $0.73$ & $0.0185$ & AGSS$09$\\
Target 9& $2.1$ & $1.18$ & $29.33$ & $262.3$& FreeEOS & OPLIB & $1.9$ & Thoul & $0.15$ & $0.73$ & $0.016$ & GN$93$\\ 
Target 10& $3.0$ & $0.38$ & $35.38$& $329.0$ & CEFF & OPAS & $1.5$& $/$& $0.1$& $0.72$ & $0.016$ & AGSS$09$\\
Target 11& $1.6$ & $2.35$ & $14.07$ & $80.6$ & FreeEOS & OPAL & $2.0$ & Paquette &$0.00$ & $0.71$ & $0.020$ &GN$93$\\ 
\hline
\end{tabular}
}
References: $^{1}$ \citet{Irwin}, $^{2}$ \citet{CEFF},$^{3}$ \citet{AGSS}, $^{4}$ \citet{GN93},$^{5}$ \citet{OPAL}, $^{6}$ \citet{Colgan},$^{7}$ \citet{Mondet}, $^{8}$ \citet{Thoul},$^{9}$ \citet{Paquette}. A ``$/$'' in the ``Diffusion'' column indicates that the effects of microscopic diffusion are not included in the model.
\end{table*}

These targets have been computed using the Li\`ege stellar evolution code (CLES \citet{ScuflaireCles}) and their frequencies have been computed using the Li\`ege oscillation code (LOSC \citet{ScuflaireLosc}). The formalism used for convection is that of the classical mixing-length theory \citep{Bohm} and overshooting, when applied, is implemented in the form of an instantaneous mixing. The temperature gradient in the overshooting region is forced to be adiabatic. Target models from Table \ref{tabTargets} and reference models from Table \ref{tabRefs} also included opacities at low temperature from \citet{Ferguson} and the effects of conductivity from \citet{Potekhin} and \citet{Cassisi}. The nuclear reaction rates we used are from \citet{Adelberger}. 

We used only low order radial modes in our study, with $n=1$ up to $n=15$. Tests were also carried out by reducing the number of observed frequencies  down to $10$ or $8$ modes. In section \ref{sec:SurfEff}, we also analyse how the results vary when the set of modes is changed to higher order, for which surface effects are much larger. For each target, the uncertainties of the frequency values were taken to be $0.03\mu \mathrm{Hz}$, similarly to the average error bars expected from datasets of the \textit{Kepler} mission. However, as we will see in the next section, most of the limitations of the determination does not come from the precision of the data, but from the low number of frequencies and systematic effects such as surface effects or the non-verification of the integral linear relations between relative frequency differences and corrections of thermodynamic quantities \citep{Dziembowski}.

\subsection{Calibrations using effective temperature and luminosity}\label{sec:TeffL}
First, we carried out inversions after having calibrated the models based solely on their effective temperatures $(T_{\mathrm{eff}})$ and luminosities $(L)$, using different physical ingredients than those used to build the $11$ targets of Table \ref{tabTargets}. In Table \ref{tabRefs}, we describe the properties of these reference models. For some of the targets, the calibration was pushed so that both reference model and targets had nearly exactly the same $T_{eff}$ and $L$, which means in turn that they have the same radius. However, biases in the mass of the models (as can be seen when comparing Tables \ref{tabRefs} and \ref{tabTargets}) ensure that the mean density is not the same for both target and reference models. Moreover, strong changes in the physical ingredients have been applied such that the differences observed in individual frequencies are not only due to a mean density mismatch, but also to an inaccurate depiction of the internal structure of the targets by their reference model. For some of the targets, we also computed inverted values using neighbouring models in the evolutionary sequence of the calibration, to demonstrate that the inversion was still efficient despite a radius mismatch\footnote{A fact that has already been demonstrated in the test cases of \citet{Reese} and \citet{Buldgen}.}. Other tests, not presented here, were also carried out using the FST formulation of convection \citep{Canuto} and led to similar results. 

\begin{table*}
\caption{Properties of the reference models used for the inversion.}
\label{tabRefs}
  \centering
    \resizebox{\linewidth}{!}{%
\begin{tabular}{r | c | c | c | c | c | c | c | c | c | c | c | c }
\hline
& Mass (M$_{\odot}$)& Age (Gy)& $R$ (R$_{\odot}$)& $L$ (L$_{\odot}$)& EOS& Opacities& $\alpha_{\mathrm{MLT}}$& Diffusion& $\alpha_{\mathrm{Ov}}$&$X_{0}$&$Z_{0}$ & Mixture  \\
\hline 
Reference 1& $1.3$ & $4.66$ & $9.00$ & $36.30$ & FreeEOS & OPAL & $2.000$ & $/$& $0.0$ & $0.71$ & $0.0155$ & AGSS$09$\\ 
Reference 2& $1.35$ & $6.30$ & $6.71$ & $17.44$ & FreeEOS & OPLIB & $1.805$ & $/$ & $0.1$ & $0.72$ & $0.028$ & AGSS$09$\\ 
Reference 3& $1.08$& $10.00$ & $14.49$ & $71.98$ & FreeEOS & OPLIB & $2.150$ & Thoul& $0.0$ & $0.70$ & $0.022$ &AGSS$09$\\ 
Reference 4& $1.18$ & $5.26$ & $29.2$ & $272.0$ & FreeEOS & OPLIB & $2.340$ & $/$& $0.1$ & $0.69$ &$0.014$ &AGSS$09$\\ 
Reference 5& $1.05$ & $11.4$ & $33.11$ & $195.4$ & FreeEOS & OPLIB & $1.600$ & $/$& $0.0$ & $0.69$ & $0.023$ &AGSS$09$ \\ 
Reference 6& $1.05$ & $9.30$ & $48.21$ & $435.08$ & FreeEOS & OPLIB & $1.900$ & $/$ & $0.15$ & $073$ & $0.022$ &GN$93$\\
Reference 7& $1.35$ & $6.69$ & $ 15.53$ & $81.88$ & CEFF & OPLIB & $1.900$ & Thoul & $0.00$ & $0.69$& $0.018$& AGSS$09$\\
Reference 8& $2.2$ & $0.82$ & $35.98$ & $350.7$ & CEFF & OPAL & $1.850$ & $/$ & $0.00$ & $0.70$ & $0.0185$ & AGSS$09$\\
Reference 9& $1.9$ & $1.32$ & $28.54$ & $248.3$& CEFF & OPLIB & $1.850$ & $/$ & $0.1$ & $0.72$ & $0.017$ & GN$93$\\ 
Reference 10& $2.8$ & $0.62$ & $35.10$& $323.2$ & CEFF & OPAL & $1.350$& $/$& $0.0$& $0.70$ & $0.013$ & AGSS$09$\\
Reference 11& $1.9$ & $1.23$ & $14.16$ & $81.7$ & FreeEOS & OPAL & $1.800$ & $/$ & $0.72$ & $0.016$ & &GN$93$\\ 
\hline
\end{tabular}
}
\end{table*}

\begin{figure*}
	\centering
		\includegraphics[width=13.5cm]{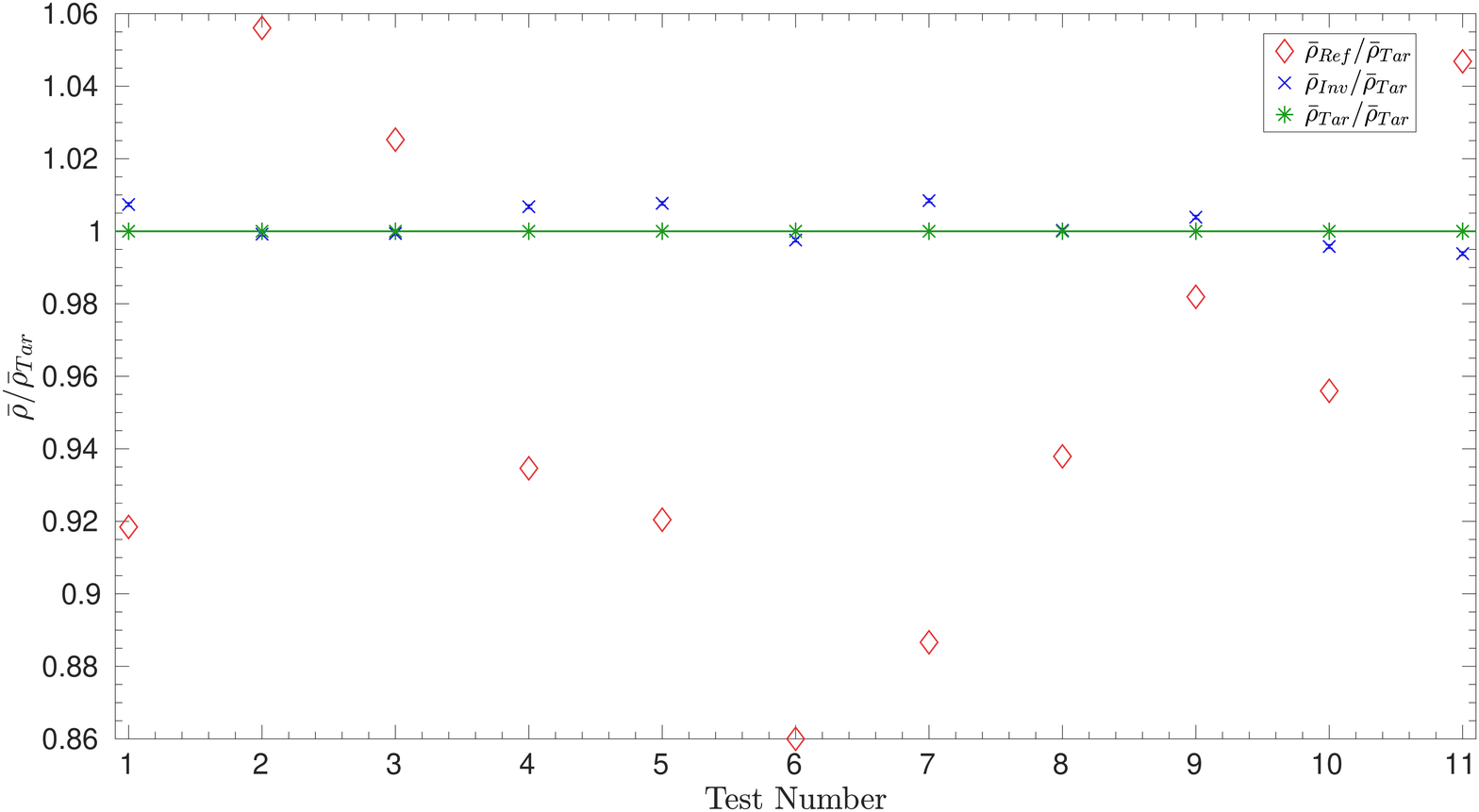}
	\caption{Mean density inversion results for exercises between the targets of Table \ref{tabTargets} and the reference models from Table \ref{tabRefs}.}
		\label{figInvResTL}
\end{figure*}

Inversion results are illustrated in Figure \ref{figInvResTL} for each pair of models, where we can see that for each test case, the inversion provides an estimate of the mean density within $1$ per cent of the real value of the artificial target. The reliability of the method is confirmed from the analysis of the value of the errors on the cross-term and averaging kernels, which remained small and of the same order of magnitude as what was observed in \citet{Reese} for main-sequence stars. Tests were also conducted with the differential formulation based on the linear fit of the average large frequency separation as in \citet{Reese} and \citet{Buldgen}, as well as the so-called KBCD approach of \citet{Reese}, based on the \citet{Kjeldsen} surface correction law. These methods are formally quite similar to the inversion, as they attempt to relate relative frequency differences to the mean density of a given target. However, the first method is based on the fact that the average large frequency separation is simply a frequency combination. From there, a differential form can be derived when comparing an observed target to a reference model and used to obtain a corrected mean density value. The KBCD approach is based on Eq. $6$ of \citet{Kjeldsen}, from which a differential form based on individual frequencies can also be derived and used to define a corrected mean density value from comparisons between observed data and a reference model. We refer the reader to \citet{Reese} for more details and the mathematical expressions associated with these methods.

Both these methods showed unstable behaviours, in the sense that they could sometimes provide results of similar quality than those of the SOLA inversion, and sometimes provide much less accurate results, with differences to the target value of more than $4$ per cent. From an in-depth investigation, we can conclude that their accuracy, when good, is actually due to a systematic compensation of their various error contributions. This result had also already been observed for main-sequence stars \citep{Reese,Buldgen}. Therefore, it is not surprising to observe a similar behaviour in these test cases using only radial modes.

\begin{figure*}
	\centering
		\includegraphics[width=13.5cm]{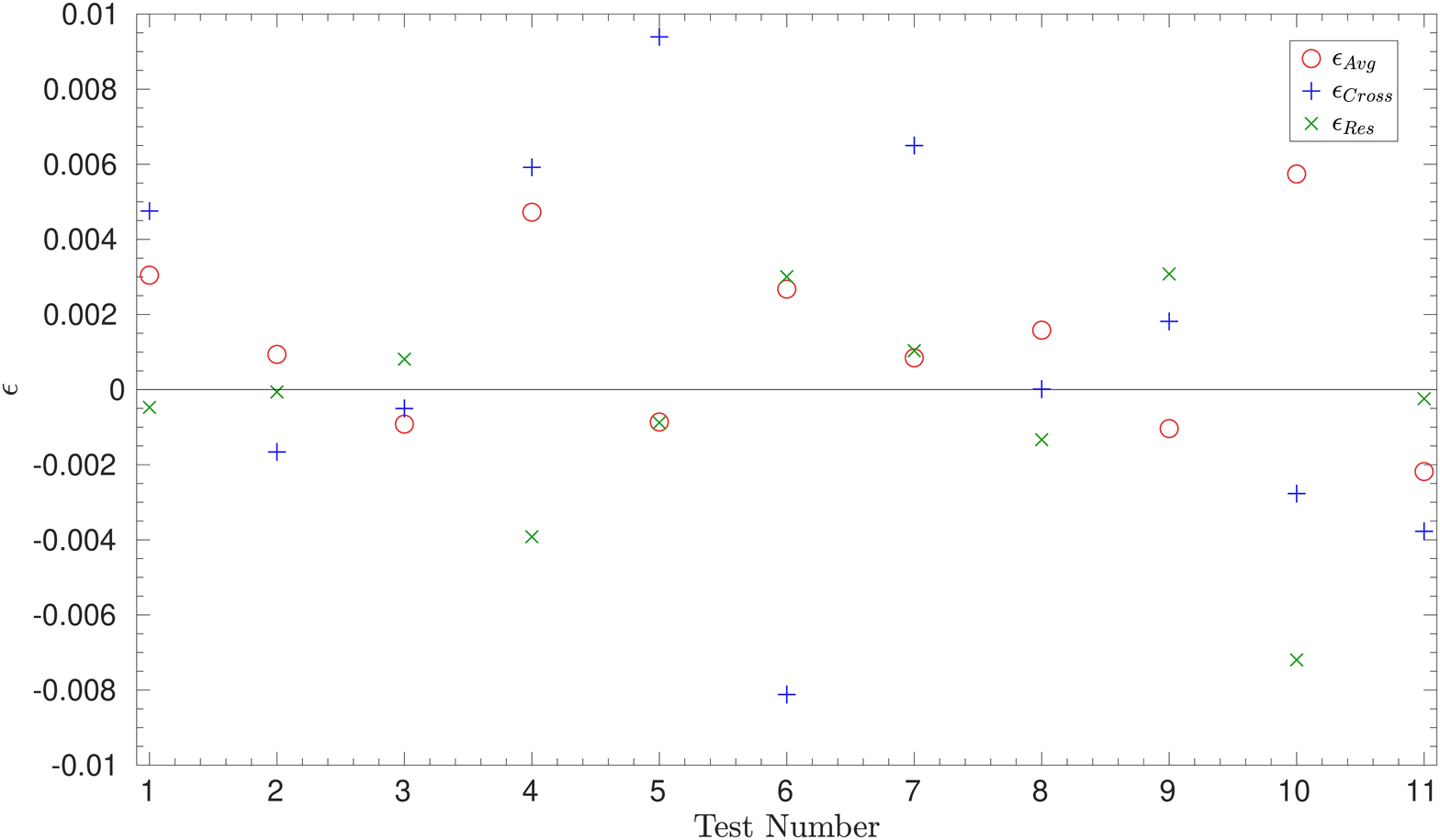}
	\caption{Error contributions $\epsilon_{Avg}$, $\epsilon_{Cross}$ and $\epsilon_{Res}$ as defined in Eqs. \ref{eqAvgErr}, \ref{eqCrossErr} and \ref{eqResErr} for each pair of target and reference models from Tables \ref{tabTargets} and \ref{tabRefs}.}
		\label{figErrorTL}
\end{figure*}

In Figure \ref{figErrorTL}, we illustrate the error contributions as defined in Eqs. \ref{eqAvgErr}, \ref{eqCrossErr} and \ref{eqResErr}. The first striking difference in the error contribution is the value of the cross-term error. In \citet{Reese}, one can see that the cross-term error associated with $\Gamma_{1}$ is very often one order of magnitude, if not more, lower than the averaging kernel and the residual error contribution.

\begin{figure}
	\centering
		\includegraphics[width=8.5cm]{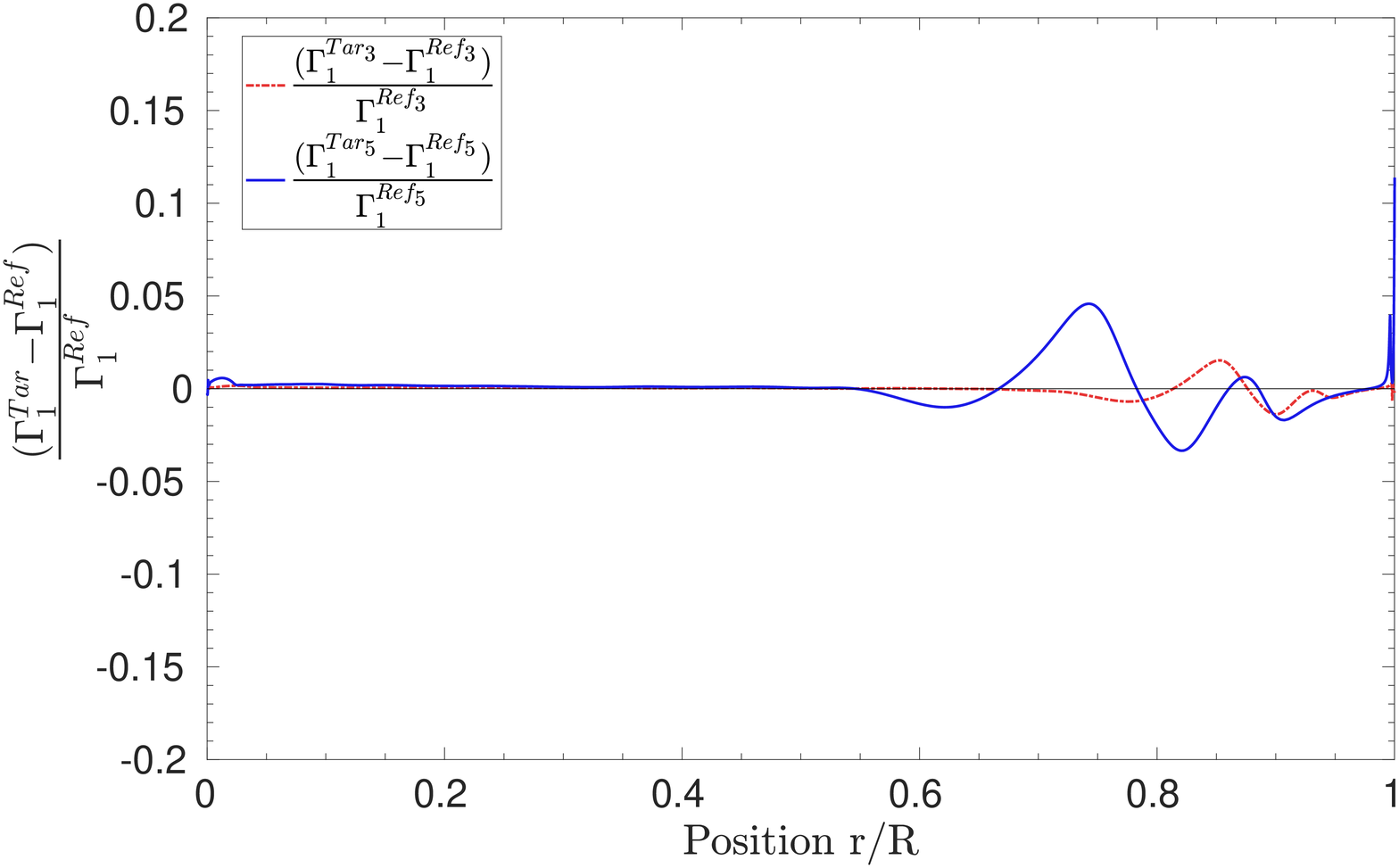}
	\caption{Relative differences in adiabatic exponent $\Gamma_{1}$ for the third and fith test cases, illustrating the impact of $\Gamma_{1}$ on the cross-term contribution. The relative differences in adiabatic exponent are here $\frac{\delta \Gamma_{1}}{\Gamma_{1}}=\frac{\Gamma^{\mathrm{Tar}}_{1}-\Gamma^{\mathrm{Ref}}_{1}}{\Gamma^{\mathrm{Ref}}_{1}}$, where ``Tar'' refers to the artificial target and ``Ref'' to the associated reference model.}
		\label{figDifGamma1}
\end{figure}

In these test cases, we can see that the cross-term error can sometimes become much more significant and even the dominant source of error. The most striking cases being Targets $5$, $6$ and $7$ in Figure \ref{figErrorTL}, where the cross-term error clearly dominates all contributions. This is a consequence of specific aspects of the inversions considered here. The very low number of modes implies that the damping of the cross-term will be very difficult. In turn, this implies that keeping a low cross-term contribution can only be achieved if the hypothesis that $\frac{\delta \Gamma_{1}}{\Gamma_{1}}$ is very small at all depths is satisfied. This is, however, not always the case and is a consequence of the large radial extent of the helium ionization zones. Consequently, the differences in $\Gamma_{1}$ due to mismatches in the helium abundance will have a larger impact on the frequencies than in main-sequence stars, where their very narrow width implies that they do not have a strong contribution on the cross-term integral. In addition to this effect, the density values, and in turn the averaging kernel term in this inversion, has a much lower value than on the main-sequence, as a consequence of the expansion of the star. Looking back at Targets $5$, $6$ and $7$ we find that these are the targets for which the differences both in mass and helium abundance are the largest of all our sample. These differences naturally implied large values for $\frac{\delta \Gamma_{1}}{\Gamma_{1}}$ between these models and thus a higher cross-term contribution. We illustrate this effect by plotting in Figure \ref{figDifGamma1} a comparison between the differences in $\Gamma_{1}$ for Target $3$, which has a low cross-term contribution and Target $5$ for which $\epsilon_{Cross}$ largely dominates. These changes also imply that the trade-off parameters \citep[see][for a thorough discussion of the trade-off problem in the context of Geophysics]{Backus} have to be properly adjusted depending on the proximity of the model with its target. In these test cases, large variations of the helium abundance were observed and had to be damped by increasing the $\beta$ parameter in the SOLA cost function. 

From Figure \ref{figErrorTL}, it is however clear that the accuracy of the inversion does not stem from large compensation effects between the various error contributions. Such a compensation would be seen if for example a large positive value of a few per cent was found for example for $\epsilon_{\mathrm{Avg}}$ in Figure \ref{figErrorTL} and a large negative value of the same order of magnitude was found for $\epsilon_{\mathrm{Cross}}$ in the same inversion. This would imply that the agreement between the inverted value and the target value is purely fortuitous and not representative of the real limitations of the inversion.

In Figure \ref{figErrorTL}, we see that for most of the cases, the sum of the modulus of all the error contributions remains within $1$ per cent and never exceeds $1.5$ per cent for the remaining cases. As a comparison, the usual scaling law using the large frequency separation and a solar reference showed at best differences of $1.3$ per cent for Target $3$, where the inversion shows differences of less than $0.1$ per cent with the observed value. For the other targets, the accuracy of the scaling laws could be worse than $30$ per cent, especially for the more massive targets. Although one could apply corrections to the scaling laws to improve their accuracy, it is clear that the inversions have the advantage of using in the most optimal way all the seismic information of the frequency spectrum. This will be further illustrated on real data in section \ref{sec:ObsBin}. Similarly, using the differential form of the $< \Delta \nu >$ scaling law defined in \citet{Reese}, who used another reference than the Sun, we showed that this formulation relied, as for main-sequence stars, on a compensation of its intrinsic errors. This implies a much lower robustness of the method, especially if only a few modes of low radial order are observed. One should also note that the test cases presented here did not use any seismic constraints to define the reference model for the inversion. Hence, the frequency differences between target and reference could sometimes be very large, which can induce a lower stability and reliability of the inversion. In practice, seismic constraints should be used to define the reference models to ensure an optimal result. The effect of seismic constraints will be presented and discussed in the next section. 

\subsection{Calibrations using seismic constraints}\label{sec:SismoCons}

In section \ref{sec:TeffL}, we only used classical parameters to define our reference models, which is far from what would be done for stars for which individual frequencies have been determined. In such cases, seismic modelling would be favoured, at least by using global parameters such as the average large frequency separation and the frequency at maximum power, $\nu_{Max}$. If individual frequencies are observed, one might even want to fit directly these constraints to extract as much seismic information as possible. While this seems an efficient approach, it should be kept in mind that directly fitting the frequencies can lead to underestimated uncertainties and strongly depends on the surface effects corrections\footnote{This is particularly true for the main-sequence stars for which a large number of frequencies are observed with \textit{Kepler}.}.

Both approaches were tested for five additional artificial targets, whose properties are given in Table \ref{tabSismTargets}. The dataset used for both the forward modelling and the inversions contained $9$ radial modes, from $n=5$ to $n=13$. The reference models were obtained through forward modelling using the AIMS pipeline. The underlying grid of models used for the fitting was built using the Li\`ege stellar evolution code (CLES) with a mass range in solar masses between $\left[0.75,2.25 \right]$ with a step of $0.02$ and a $\left[ \mathrm{Fe/H} \right]$ range between $\left[-0.75, 0.25\right]$ with a step of $0.25$. The mixing-length parameter value is fixed by a solar calibration at a value of $1.691$. The effects of microscopic diffusion are not taken into account in the models. Moreover, the grid is built with the GN$93$ abundance tables \citep{GN93}, the FreeEOS equation of state \citep{Irwin} and the OPAL opacities \citep{OPAL}, while the nuclear reaction rates from \citet{Adelberger} are used.

The results of this modelling are shown in Table \ref{tabSismRef}. We denoted the reference models computed from the fitting of individual frequencies as $1.1$, $2.1$ and so on, whereas models denoted as $1.2$, $2.2$, etc were built using the global seismic indicators, namely the average large frequency separation, $< \Delta \nu >$, and the frequency of maximum power, $\nu_{Max}$. Classical constraints such as $\left[\mathrm{Fe/H}\right]$ and $T_{eff}$ were also used, where uncertainties of $0.1$ dex and $80$K were respectively considered. No surface correction was used when carrying out the fits, as only models using similar atmospheric models and adiabatic frequencies were used here. These effects will be discussed separately, in section \ref{sec:SurfEff}.

\begin{table*}
\caption{Properties of the target models used for the inversion and AIMS modelling.}
\label{tabSismTargets}
  \centering
    \resizebox{\linewidth}{!}{%
\begin{tabular}{r | c | c | c | c | c | c | c | c | c | c | c | c }
\hline
& Mass (M$_{\odot}$)& Age (Gy)& $R$ (R$_{\odot}$)& $L$ (L$_{\odot}$)& EOS& Opacities& $\alpha_{\mathrm{MLT}}$& Diffusion& $\alpha_{\mathrm{Ov}}$&$X_{0}$&$Z_{0}$ & Mixture  \\
\hline 
Target 1& $1.40$ & $3.09$ & $2.77$ & $6.44$ & FreeEOS & OPAL & $1.8$ & $/$& $0.0$ & $0.72$ & $0.0135$ & AGSS$09$\\ 
Target 2& $1.05$ & $9.85$ & $4.01$ & $6.50$ & FreeEOS & OPLIB & $2.0$ & Paquette & $0.0$ & $0.71$ & $0.0200$ & GN$93$\\ 
Target 3& $1.15$& $11.61$ & $3.67$ & $4.67$ & CEFF & OPLIB & $1.5$ & Paquette& $0.15$ & $0.72$ & $0.0300$ &AGSS$09$\\ 
Target 4& $1.05$ & $7.94$ & $2.47$ & $3.47$ & FreeEOS & OPLIB & $1.8$ & Thoul& $0.2$ & $0.73$ &$0.0100$ &AGSS$09$\\ 
Target 5& $1.20$ & $8.31$ & $2.61$ & $3.64$ & FreeEOS & OPAL & $2.0$ & $/$& $0.15$ & $0.71$ & $0.0300$ &GN$93$ \\ 
\hline
\end{tabular}
}
\end{table*}

\begin{figure}
	\centering
		\includegraphics[width=7cm]{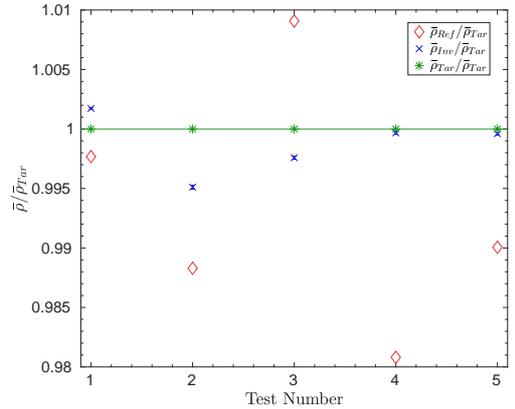}
	\caption{Inversion results for the artificial targets $1$ to $5$, using the reference models $1.2$ to $5.2$. The abscissa refers to the target number.}
		\label{figInvSismo}
\end{figure}

The first conclusion that can be drawn is that the fit of the mean density is very good using both global seismic constraints and individual frequencies. More specifically, the fit of the mean density is exact to numerical precision when the individual modes are used and in fact, since almost every frequency is fitted within its error bars, the inversion process is in this case useless. Indeed, by definition, the inversion, based on the recombination of individual frequency differences, will not bring any additional information. This is the reason why the Reference $1.1$, $2.1$, $3.1$, $4.1$ and $5.1$ are not illustrated in Figure \ref{figInvSismo}. However, we can still see that the masses and radii of the artificial targets are not always properly reproduced. This is due to the incorrect reproduction of the helium abundance, since the grid used for the AIMS modelling was built using a fixed enrichment law in heavy elements abundances and the targets were built without following any such approach. This implies that the determination of stellar mass on the RGB still strongly relies on accurate determinations of stellar luminosities and/or radii, as provided by the Gaia mission.

\begin{table*}
\caption{Properties of the reference models used for the inversion and AIMS modelling.}
\label{tabSismRef}
  \centering
    \resizebox{\linewidth}{!}{%
\begin{tabular}{r | c | c | c | c | c | c | c | c | c | c | c | c }
\hline
& Mass (M$_{\odot}$)& Age (Gy)& $R$ (R$_{\odot}$)& $L$ (L$_{\odot}$)& EOS& Opacities& $\alpha_{\mathrm{MLT}}$& Diffusion& $\alpha_{\mathrm{Ov}}$&$X_{0}$&$Z_{0}$ & Mixture  \\
\hline 
Reference 1.1& $1.402$ & $3.15$ & $2.77$ & $6.49$ & FreeEOS & OPAL & $1.691$ & $/$& $0.05$ & $0.719$ & $0.0157$ & GN$93$\\ 
Reference 1.2& $1.423$ & $3.087$ & $2.78$ & $6.67$ & FreeEOS & OPAL & $1.691$ & $/$ & $0.05$ & $0.716$ & $0.0173$ & GN$93$\\ 
Reference 2.1& $1.113$& $7.50$ & $4.1$ & $7.50$ & FreeEOS & OPAL & $1.691$ & $/$& $0.05$ & $0.721$ & $0.0134$ &GN$93$\\ 
Reference 2.2& $1.153$ & $6.92$ & $8.25$ & $4.16$ & FreeEOS & OPAL & $1.691$ & $/$& $0.05$ & $0.718$ &$0.0161$ &GN$93$\\ 
Reference 3.1& $1.082$ & $11.30$ & $3.60$ & $5.41$ & FreeEOS & OPAL & $1.691$ & $/$& $0.05$ & $0.679$ & $0.0331$ &GN$93$ \\ 
Reference 3.2& $1.251$ & $6.38$ & $3.77$ & $6.39$ & FreeEOS & OPAL & $1.691$ & $/$& $0.05$ & $0.691$ & $0.0300$ &GN$93$ \\
Reference 4.1& $1.041$ & $9.73$ & $2.46$ & $3.23$ & FreeEOS & OPAL & $1.691$ & $/$& $0.05$ & $0.718$ & $0.0149$ &GN$93$ \\
Reference 4.2& $1.212$ & $5.80$ & $2.61$ & $3.72$ & FreeEOS & OPAL & $1.691$ & $/$& $0.05$ & $0.711$ & $0.0194$ &GN$93$ \\
Reference 5.1& $1.157$ & $8.43$ & $2.58$ & $3.13$ & FreeEOS & OPAL & $1.691$ & $/$& $0.05$ & $0.679$ & $0.0331$ &GN$93$ \\
Reference 5.2& $1.329$ & $4.75$ & $2.71$ & $3.89$ & FreeEOS & OPAL & $1.691$ & $/$& $0.05$ & $0.695$ & $0.0274$ &GN$93$ \\ 
\hline
\end{tabular}
}
\end{table*}

Moreover, we will see in section \ref{sec:ObsBin} that mean density inversions can still be useful in real observed cases. First, because it is in practice nearly impossible to fit every individual mode for real data and the results still depend in any case from the adopted surface corrections. Second, because the inversion can provide a useful verification step to the forward modelling process and provide a mean density value which can then be directly introduced in the cost function of second forward modelling step.

\begin{figure}
	\centering
		\includegraphics[width=8cm]{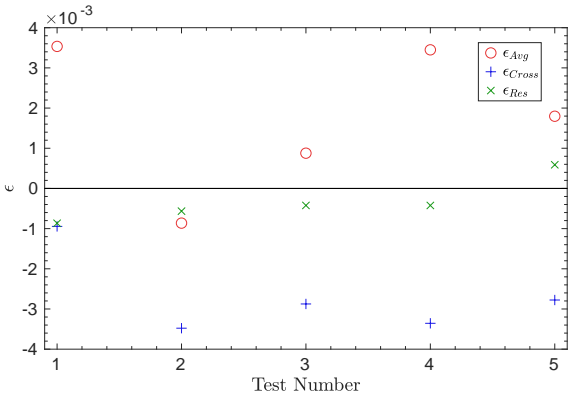}
	\caption{Error contributions $\epsilon_{Avg}$, $\epsilon_{Cross}$ and $\epsilon_{Res}$ as defined in Eqs. \ref{eqAvgErr}, \ref{eqCrossErr} and \ref{eqResErr} for the inversion exercises using targets $1$ to $5$ and reference models $1.2$ to $5.2$.}
		\label{figErrorSismo}
\end{figure}

From Table \ref{tabSismRef}, we can see that the fits using global seismic parameters can lead to less accurate results, even for the mean density. In these cases, since the individual frequencies were not reproduced within their uncertainties, computing the seismic inversions could provide an improvement of the mean density value. These results are illustrated in Figure \ref{figInvSismo} and the associated error contributions are illustrated in Figure \ref{figErrorSismo}.

From Figure \ref{figErrorSismo}, we see that the SOLA method is able to provide a better determination of the mean density without large compensation of its errors. A certain degree of compensation can be observed, but the modulus of each error contribution remains well within $1$ per cent, ensuring that the inversion still provides additional information after the forward modelling process. This confirms that the use of seismic inversions offers a gain in accuracy and tighter constraints on the mean density. The inverted value can than be used directly as constraint in a second run of seismic forward modelling.

\subsection{Red clump stars}

In addition to RGB targets, two red clump models of the same evolutionary sequence were also used to test the inversion techniques. The properties of these targets are given in Table \ref{tabRCTarRef}. We first made attempts to carry out inversions for these artificial targets using a clump and asymptotic giant branch models as references. These references were built using slightly different physical ingredients. Reference $1$ was calibrated to approximately reproduce both $T_{eff}$ and $\nu_{Max}$ of the targets. References $2$ and $3$ are two additional models tested with both Target $1$ and $2$. They were just taken as additional test cases to see how far the inversion could be pushed once it was established that the star was in the helium burning phase. Both reference models and artificial targets were computed using the MESA evolution code \citep{MESAPAX} and their characteristics are summarized in Table \ref{tabRCTarRef}. Again, we used $9$ radial oscillations as dataset for the inversion, namely the modes with $n=5$ to $n=13$.

\begin{table*}
\caption{Properties of the clump targets and references.}
\label{tabRCTarRef}
  \centering
    \resizebox{\linewidth}{!}{%
\begin{tabular}{r | c | c | c | c | c | c | c | c | c | c | c | c }
\hline
& Mass (M$_{\odot}$)& Age (Gy)& $R$ (R$_{\odot}$)& $L$ (L$_{\odot}$)& EOS& Opacities& $\alpha_{\mathrm{MLT}}$& Diffusion& $\alpha_{\mathrm{Ov}}$&$X_{0}$&$Z_{0}$ & Mixture  \\
\hline 
Target 1 & $1.00$ & $12.30$ & $10.38$ & $48.88$ & OPAL & OPAL & $1.966$ & $/$& $0.2$ & $0.7163$ & $0.0176$ & GN$93$\\ 
Target 2 & $1.00$ & $12.32$ & $10.86$ & $52.31$ & OPAL & OPAL & $1.966$ & $/$ & $0.2$ & $0.7163$ & $0.0176$ & GN$93$\\ 
Reference 1 & $1.15$& $8.76$ & $11.21$ & $51.37$ & OPAL & OPAL & $1.966$ & $/$& $0.2$ & $0.696$ & $0.0276$ &GN$93$\\ 
Reference 2 & $1.15$ & $8.845$ & $12.83$ & $64.89$ & OPAL & OPAL & $1.966$ & $/$& $0.2$ & $0.718$ &$0.0161$ &GN$93$\\ 
Reference 3 & $1.15$ & $8.847$ & $13.44$ & $69.52$ & OPAL & OPAL & $1.966$ & $/$& $0.2$ & $0.718$ &$0.0161$ &GN$93$\\ 
Reference 4 & $0.91$ & $22.03$ & $9.85$ & $29.05$ & FreeEOS & OPAL & $1.691$ & $/$& $0.05$ & $0.690$ &$0.0300$ &GN$93$\\ 
Reference 5 & $0.97$ & $17.27$ & $10.47$ & $32.90$ & FreeEOS & OPAL & $1.691$ & $/$& $0.05$ & $0.690$ &$0.0300$ &GN$93$\\ 
\hline
\end{tabular}
}
\end{table*}

\begin{figure*}
	\centering
		\includegraphics[width=13.5cm]{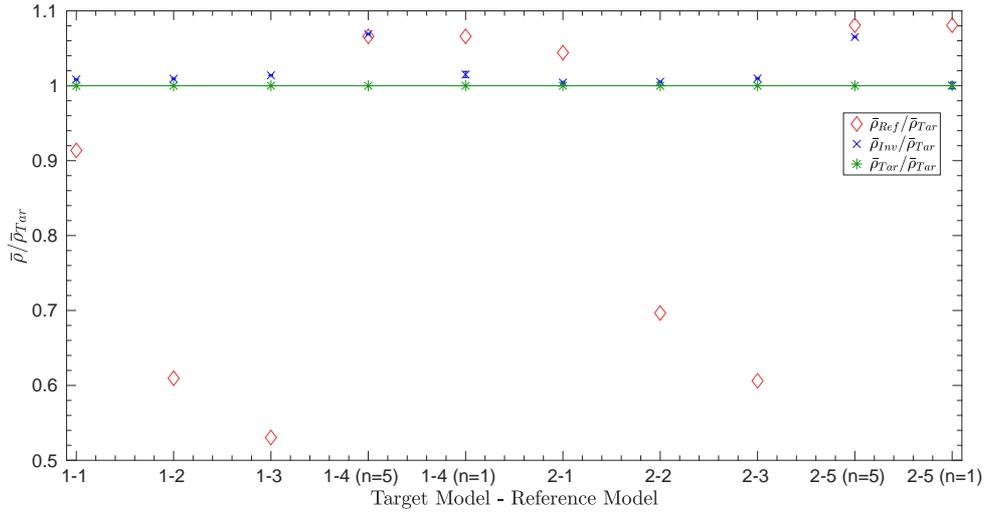}
	\caption{Inversion results for the clump models $1$ and $2$. The abscissa refers to the pair of target and reference models used for the inversion.}
		\label{figInvClump}
\end{figure*}

\begin{figure*}
	\centering
		\includegraphics[width=13.5cm]{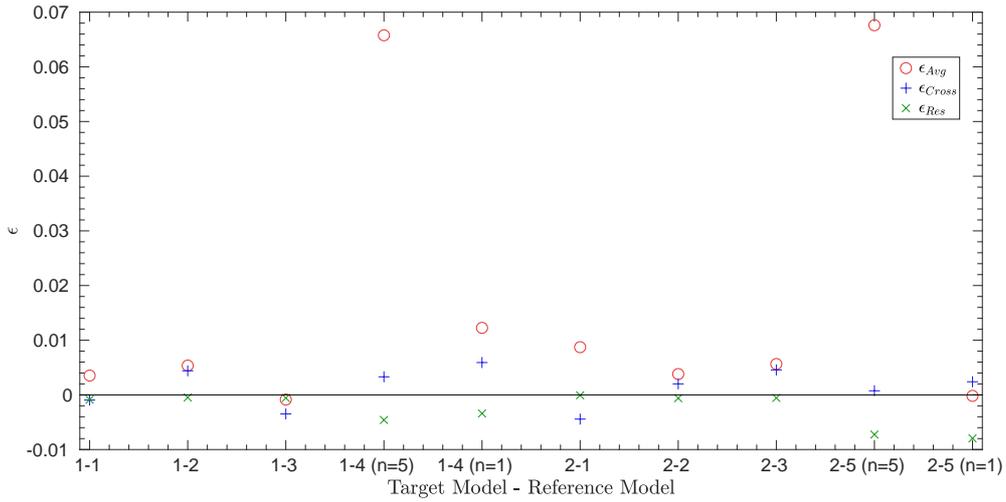}
	\caption{Error contributions $\epsilon_{Avg}$, $\epsilon_{Cross}$ and $\epsilon_{Res}$ as defined in Eqs. \ref{eqAvgErr}, \ref{eqCrossErr} and \ref{eqResErr} for the inversion exercises using clump stars as artificial targets. The abscissa refers to the pair of target and reference models used for the inversion.}
		\label{figErrorContribClump}
\end{figure*}

The results of these exercises are illustrated in Figure \ref{figInvClump}, and the error contributions to the inversions are illustrated in Figure \ref{figErrorContribClump}. From these tests cases, it seems that mean density inversions can also be performed for clump stars even if only a small number of radial frequencies are available. The typical errors of these inversions is then of around $1$ per cent, which is much better than what can be achieved using the average large frequency separation.

While these tests demonstrate the efficiency of the method, they also rely on a fundamental hypothesis, which is that one was able to determine that the target belonged to the clump and not the RGB. For stars with an observed period spacing, there is no ambiguity possible, but higher on the asymptotic giant branch and even for some clump stars, no period spacing can be determined. For these specific cases, confusion can remain and the reference model can be very far from its actual target.

To simulate such effects, we used the AIMS software to fit the individual frequencies of the clump models using a grid of RGB models only. As a result, we got reference models very far from the targets but which actually reproduced quite well (although not as well as for the tests on the RGB) the target frequencies. These models are denoted Reference $4$ and $5$ and display for example non-physical ages and are completely off in terms of luminosity and $T_{eff}$, which were not used in the fits. Therefore, in a real case, these models could clearly be rejected, but to further test and break the inversion, we still kept them and attempted to determine the mean density of both Target $1$ using Reference $4$ and Target $2$ using Reference $5$.

These test cases are very enlightening as neither the fitting process using AIMS nor the inversion could provide a reliable estimate of the mean density. This is illustrated in Figures \ref{figInvClump} and \ref{figErrorContribClump}. In Figure \ref{figErrorContribClump}, we can see that the SOLA method is unable to provide an accurate fit of the target function and hence a reliable mean density value. The only way for the inversion to work was to include the fundamental harmonics in the frequencies used for the inversion. We have added this test case in Figures \ref{figInvClump} and \ref{figErrorContribClump} to illustrate the effect of using the full set of modes.

This is not really surprising, as the fundamental mode is the one that carries most of the information on the deep layers. Its value was indeed radically different from that of the corresponding modes of the artificial target, while all other modes could be fitted quite well. This illustrates two things, first, that the fundamental radial mode can be used to distinguish between RGB and red clump stars, and second, that even if the frequencies seem well fitted, the mean-density of the star is not. This is in strong contrast with the RGB case and demonstrates again the importance of providing as reliable as possible reference models when computing linear structural inversions. Indeed, these test cases simply illustrate the well-known fact that inversions in asteroseismology are not fully model-independent and thus require a proper assessment of the model-dependency to avoid biased and hasty interpretations of their results. 

\subsection{Importance of surface effects}\label{sec:SurfEff}

As stated in section \ref{sec:inversion}, surface effects are an important source of uncertainties for seismic inversions based on the integral relations for individual frequencies. In recent years, various corrections have been proposed to take these effects into account in seismic modelling. The most recent ones, tested with patched models, are those of \citet{Ball1} and \citet{Sonoi}. In this section, we test these corrections when applied to the inversion of the mean density of RGB stars, using the patched model I of \citet{Sonoi} as a target, computed with the CESTAM stellar evolution code \citep{CESTAMI,CESTAMII}. Frequencies for this model were computed in the adiabatic approximation using the Aarhus pulsation package \citep{JCDAdipls} and the MAD non-adiabatic oscillation code \citep{Dupret,Dupret06,Grigahcene} which takes into account a time-dependent treatment of convection including the effects of turbulent pressure and variations of the convective flux due to the oscillations.

In addition to the various corrections proposed in the literature, there are multiple ways of including these corrections in the SOLA inversion. The classical approach is to mimic what has been done in helioseismology, by including an additional term in the SOLA cost function of Eq. \ref{eq:CostSOLA}, which will then have a specific form depending on the surface correction chosen. The parameters for the correction are free parameters of the inversion, to be determined alongside the structural corrections. However, this approach might be considered suboptimal in asteroseismology, where the number of individual modes is very limited and where including the surface correction might lead to low quality fits of the averaging kernels. This problem is especially true for the cases studied here, where one might have around $10$ individual modes or less to extract the structural constraints. Another approach is to correct the frequencies before carrying out the inversion, using the empirical law provided in \citet{Sonoi} or the coefficients from \citet{Ball1} and \citet{Ball2} obtained while carrying out the forward modelling\footnote{This approach might induce correlations between the frequencies. In this study, since the coefficients of the empirical law of \citep{Sonoi} are fixed from non-seismic parameters, no correlation is introduced. Similarly, the \citep{Ball1} correction has been implemented in AIMS as an additional free parameter of the forward modelling process and used to correct the theoretical frequencies, which also avoids the introduction of additional correlations.}. The inversion is then carried out assuming that the surface effects have been corrected and no surface term is then added to the SOLA cost function. In what follows, we will provide illustrations of both approaches.

\begin{figure}
	\centering
		\includegraphics[width=7.5cm]{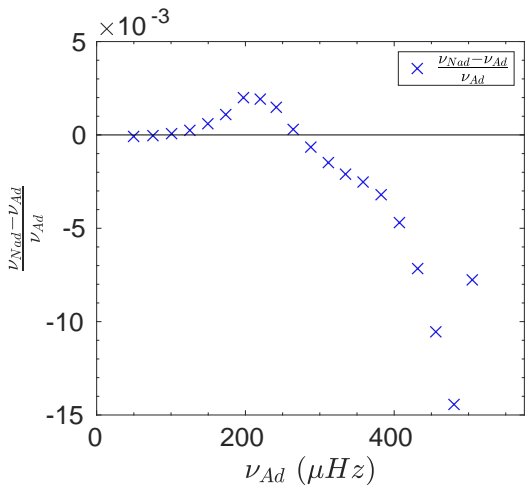}
	\caption{Relative differences between adiabatic and non-adiabatic frequencies in frequency plotted against frequency for the radial modes of the patched model considered as target for our exercise.}
		\label{figDiffFreq}
\end{figure}

Reference models were obtained with the AIMS software \citep{AIMSSoftware,ReeseHH,Rendle} using both adiabatic and non-adiabatic frequencies for the target patched model. The reference models were built using various sets of constraints, using the Li\`ege stellar evolution code \citep{ScuflaireCles} and included an Eddington grey atmosphere. A first reference model for the artificial target was determined using the effective temperature, the average large frequency separation of the radial modes determined using a linear regression and the frequency of maximum power, $\nu_{Max}$, derived from the scaling laws. However, to see the impact of the surface effects when using individual modes, we also carried out a second fit of the artificial target using the effective temperature and the individual frequencies of the radial mode as constraints. No surface correction was considered in either case, implying that both fits should be biased. We considered low order modes, from $n=1$ to $n=15$, which have low frequencies compared to $\nu_{Max}$ and thus should not display extreme changes due to the upper layers. This is confirmed in Figure \ref{figDiffFreq} where we compare the adiabatic and non-adiabatic frequencies from $n=1$ up to $n=20$. As expected, a clear trend with frequency is observed. 

The results using AIMS for the forward modelling are the following. Using the mean large frequency separation, the difference of the mean density value is of $3.6$ per cent when adiabatic frequencies are considered. If non-adiabatic frequencies are used in the modelling, then the difference between the target value and the value determined by AIMS is of $3.8$ per cent. The fact that using the non-adiabatic frequencies only induces a small change of the results is due to the properties of the dataset, which contains low frequencies for which non-adiabatic effects are small. If one uses individual frequency values as seismic constraints, the difference is reduced by a factor $3$, going down to $1.1$ per cent if adiabatic frequencies are considered and to $1.3$ per cent if non-adiabatic frequencies are used. 

\begin{figure*}
	\centering
		\includegraphics[width=13.5cm]{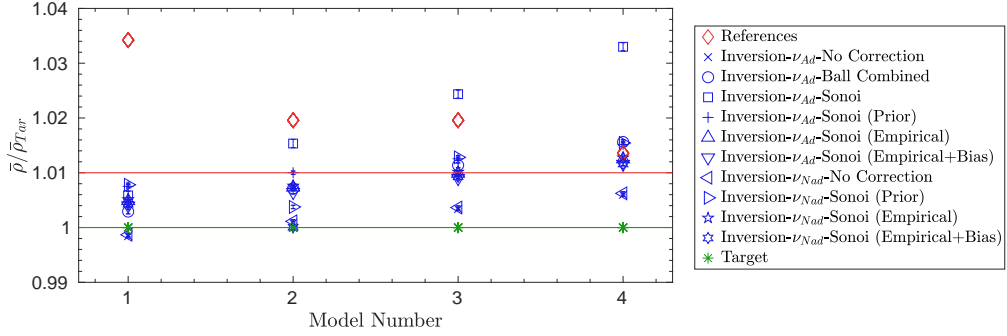}
	\caption{Inversion results using various surface effects corrections for each of the $4$ models using patched model I from \citet{Sonoi} as a target and either adiabatic or non-adiabatic frequencies.}
		\label{figInvRhoNad}
\end{figure*}

We then tested the inversion technique using $4$ reference models close to the models found with AIMS, to also assess the model dependency of the inversion technique. The results are illustrated in Figure \ref{figInvRhoNad} and the error contributions are given in Figure \ref{figErrorNoNad} for each model and most of the surface effect corrections. In Figure \ref{figInvRhoNad}, the notation ``prior'' and ``posterior'' has been adopted when the \citet{Sonoi} correction was respectively applied on the frequencies beforehand, using the coefficients directly from the paper and when the correction was applied in the SOLA cost function directly in a linearized form. The notation ``empirical'' was used to denote the case where the coefficients were derived from the empirical formula in $T_{\mathrm{eff}}$ and $\log g$ from their paper.

In Figure \ref{figErrorNoNad}, we chose not to plot the error contributions of some of the test cases to avoid redundancy.

A quick inspection of the results show that the inversion results are always well below the $3.7$ per cent agreement with the target value showed by the forward modelling process using only the average large frequency separation. However, not all results are superior to the $1.1$ per cent differences from the use of individual radial modes. To better understand what is happening here, we have to take a look at the individual error contributions and see which one is contributing to the biases in some of the inversions. A first inspection shows that the best results are often obtained when no correction is applied, except for model $2$ where compensation is found for all other inversions and the excellent agreement is thus fortuitious. This is a consequence of the low frequency modes used in the set, for which all surface effect corrections seem to be overestimated and thus bias the results, we will see how these effects change when another set of modes is used for the inversion.

A second conclusion that can be drawn is that for most cases, applying the correction within the SOLA cost function is not the best option. Only model $1$ still shows good results with this implementation of the surface effect correction. In all other cases, a significant increase in the averaging kernel error $\epsilon_{\mathrm{Avg}}$ is seen and the quality of the results is relatively poor for both the \citet{Ball2} and the \citet{Sonoi} correction laws. The increase of $\epsilon_{\mathrm{Avg}}$ is simply due to the fact that including the correction within the SOLA cost function will introduce a strong trend in the averaging kernel which depends on the form of the correction. This trend then leads to a much less accurate fit of the target function of the inversion and thus to a much less accurate result.

\begin{figure*}
	\centering
		\includegraphics[width=15.5cm]{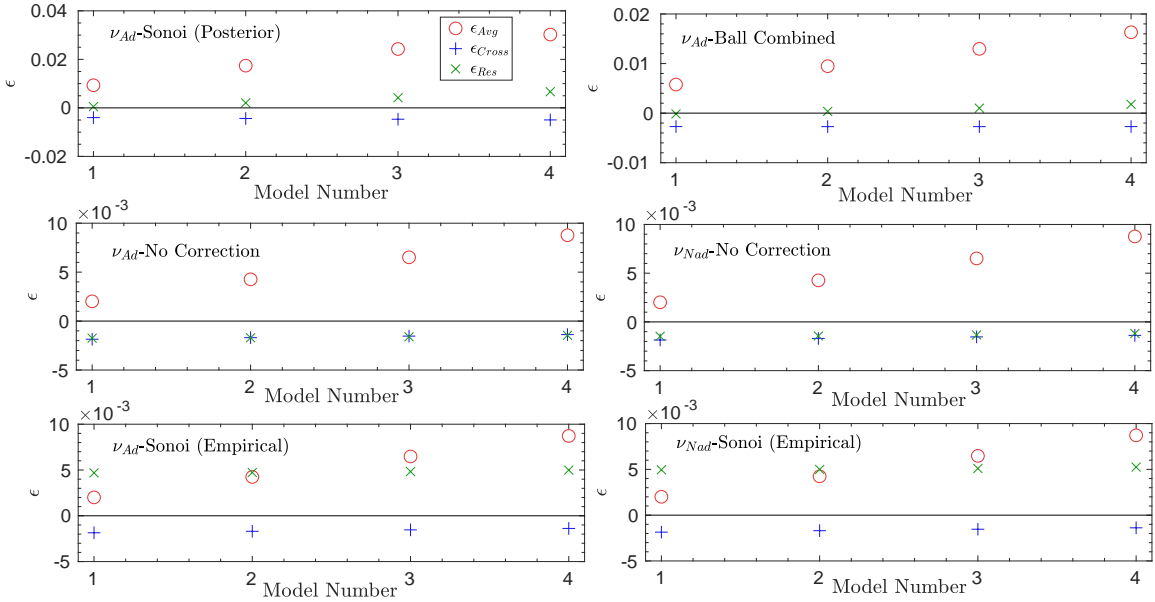}
	\caption{Error contributions $\epsilon_\mathrm{{Avg}}$, $\epsilon_{\mathrm{Cross}}$ and $\epsilon_{\mathrm{Res}}$ as defined in Eqs. \ref{eqAvgErr}, \ref{eqCrossErr} and \ref{eqResErr} for each of the $4$ models selected as a reference for the patched target for some test cases of the surface effects. Each test case has its dedicated subplot to make the analysis easier.}
		\label{figErrorNoNad}
\end{figure*}

The other approach we proposed was to introduce the surface corrections using the empirical law proposed by \citet{Sonoi}. Overall, we can see from Figure \ref{figErrorNoNad} that this approach leads to larger residual errors, hence, this method does not seem to reduce significantly the surface effect for the set of modes considered. However, it has the strong advantage of not reducing the fit of the target function. The poor performance of the surface correction laws in this particular test case is a consequence the very low frequency of the modes, which results in an overestimate of the surface corrections for the very low $n$ modes which are the most useful to the inversion. Also, it seems that biasing the correction law only slightly affects the inversion results. From Figure \ref{figInvRhoNad}, it appears that shifts of less than $0.1$ per cent of the inverted mean density values are induced by changing the $T_{\mathrm{eff}}$ and $\log g$ values in the empirical law of \citet{Sonoi}. In this particular case, we shifted the values of the quantities by $45K$ and $0.1$dex respectively. Larger shifts would induce larger deviations but they would remain negligible in the total error budget of the inverted results. From this analysis, it seems that the \citet{Ball2} combined law does the best job at reducing the residual error and thus the contribution of surface effects. However, this reduction of the residual error is made at the expense of a larger increase of the error from the fit of the averaging kernels when this approach is directly implemented in the SOLA cost function. It thus seems that the optimal choice would be to be able to apply the \citet{Ball2} correction before carrying out the inversion or defining some law with a similar form as in \citet{Sonoi} for this surface correction.

\begin{figure}
	\flushleft
		\includegraphics[width=9.4cm]{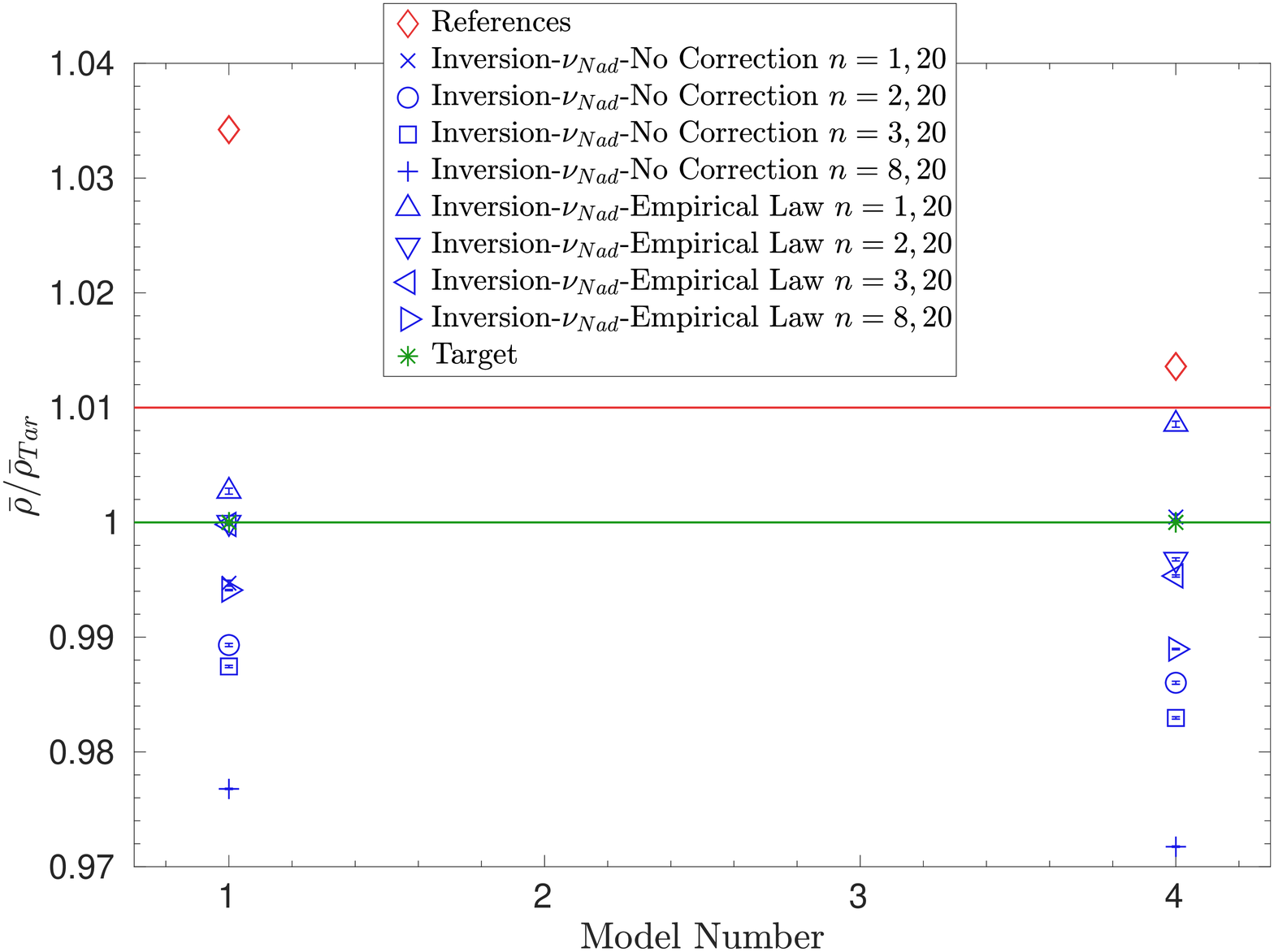}
	\caption{Inversion results for models $1$ and $4$ using patched model I from \citet{Sonoi} as target and varying the lowest $n$ in the set of modes used for the inversion.}
		\label{figInvRhoNadHighN}
\end{figure}

To further test the impact of surface effects, we take a look at the impact of changing the set of observed modes and using only modes of higher frequencies when non-adiabatic effects are taken into account. The results are illustrated in Figure \ref{figInvRhoNadHighN} for Model $1$ and $4$. For both models, we either considered not including any surface correction or using the empirical formula of \citet{Sonoi} to correct the frequencies before carrying out the inversion. We started with the full set of non-adiabatic frequencies available for the target, thus with modes of $n=1$ up to $n=20$ and proceeded to eliminate some of the lowest order modes up to $n=7$.

From Figure \ref{figInvRhoNadHighN}, one can see that the results strongly depend on the set of modes used to carry out the inversion and that, as expected, the lowest order modes are crucial to derive a good result. One could be tempted to say that for model $1$ the empirical law does an excellent job at correcting the frequencies. However, looking at Figure \ref{figErrorNoNad}, we can see that the excellent results for the empirical correction are simply due to a fortuitous compensation effect. These test cases have also revealed a fundamental limitation of the seismic determination of the mean density using the radial oscillations of red giant stars. It appears that without the fundamental mode, the accuracy of the inversion is at best of $1.5$ per cent and that if only high $n$ modes are available, the mean density cannot be determined within less than $1$ per cent to the observed value. However, if the fundamental mode is present, the accuracy of the inversion goes down to less than $1$ per cent and is not strongly affected by the surface effects. The importance of the fundamental mode can also be observed when just looking at the change of the quality of the fit of the averaging kernel in Figure \ref{figErrorNoNad}. The importance of this specific mode is in fact not a surprise, since it carries a lot of information about the mass distribution inside the star \citep{Ledoux41,Ledoux55,Ledoux55b} and thus is crucial to seismic methods.  

We can see that for both models, if the set of observed modes is reduced to $n$ higher than $7$, the quality of the inversion result is greatly reduced. This is observed whether a surface correction is applied or not, since $\epsilon_{\mathrm{Avg}}$ quickly becomes the dominant source of errors. We can also see that the residual error, $\epsilon_{\mathrm{Res}}$, used here to quantify the surface effects contribution, rises if no correction is applied. However, we have also noted that the empirical correction of \citet{Sonoi} used in this test case does not always reduce efficiently the contribution of the surface effects. The \citet{Ball2} correction could perhaps provide a better reduction of the surface effect but it should not be directly implemented in the SOLA cost function, as it then strongly reduces the quality of the fit of the target function.

\begin{figure*}
	\centering
		\includegraphics[width=15.5cm]{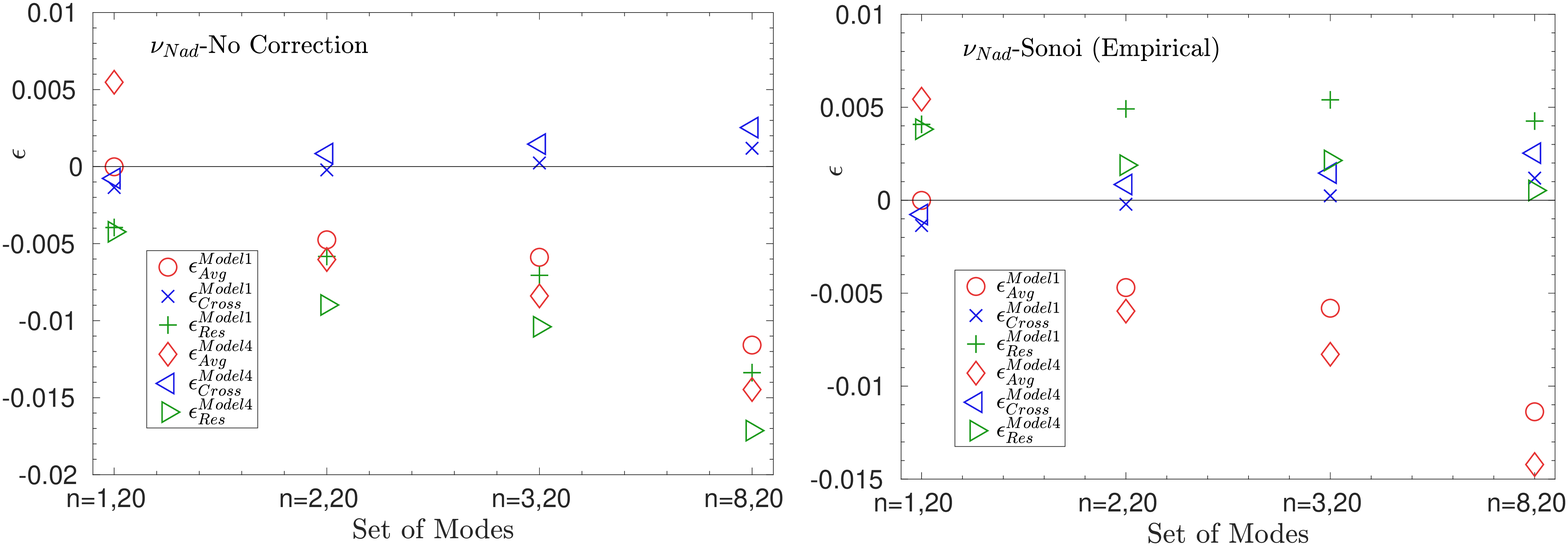}
	\caption{Error contributions $\epsilon_{Avg}$, $\epsilon_{Cross}$ and $\epsilon_{Res}$ as defined in Eqs. \ref{eqAvgErr}, \ref{eqCrossErr} and \ref{eqResErr} for models $1$ and $4$, varying the lowest $n$ in the set of observed frequencies.}
		\label{figErrorNoNad}
\end{figure*}

In addition to the seismic inversion of the mean density, we also tested the variational formulation of the scaling law for the average large frequency separation determined from a linear regression, presented in \citet{Reese}. For all test cases presented here, this approach provided worse results than the SOLA inversion. This is also seen in the AIMS fits, where using the average large frequency separation leads to a $3.7$ per cent error on the estimate of the mean density. This approach did not provide good results and was found to be very sensitive to surface effects. For example, the results significantly varied if the empirical correction of \citet{Sonoi} was biased. We also found that the quality of the results strongly varied with the set of modes. For example, if low $n$ modes are used, the impact of surface effects is strongly reduced but since the modes are far from being in the asymptotic regime, the accuracy remains very low. Results obtained from the large frequency separation improve if higher order modes are taken, but are then affected by the surface effects. In addition, we observed, as in \citet{Reese}, that the accuracy of this method relied on error compensations rather than a high-quality fit of the kernels. This implies that surface effects, non-linear behaviour, or an inadequate verification of the asymptotic relations can quickly lead to disagreements with the real value as large as $2$ or $3$ per cent. In fact, in a worst case scenario, using non-adiabatic frequencies, differences as high as $8$ per cent were observed for the mean density determined from the average large frequency separation.

Overall, it seems that using the information of the individual modes offers much better constraints, since the fits obtained with AIMS could reach an agreement with the target value of around $1$ per cent. A mean density inversion can in such cases provide additional information and reach below the $1$ per cent limit for the mode set considered here. Depending on the set of modes, the accuracy of the inversion technique will vary, but so will the accuracy of the forward modelling technique. We emphasize here that the $1$ per cent differences of the forward modelling process were obtained for a set of modes where the fundamental mode was observed. Additional tests on a larger sample of artificial targets and observed modes are thus required to provide better insights on the problem of surface effects and its impact on forward modelling approaches.

An additional comment should also be made regarding the way the surface effects are simulated in this study. First, the modelling of the non-adiabatic effects is far from perfect and problems still remain in the physical representation of the coupling between convection and oscillations, as can be seen from the differences in linewidth and amplitudes between simulated non-adiabatic spectra and observations in \citet{Grosjean}. This implies that the tests we carried out with non-adiabatic frequencies are only qualitative but can serve as a warning when using various surface correction laws which are derived from purely adiabatic computations. 

\section{Application to eclipsing binaries}\label{sec:ObsBin}

In addition to numerical exercises on artificial targets, we also tested the mean density inversions on a few eclipsing binaries observed by \textit{Kepler}, previously studied by \citet{Gaulme2016} and \citet{Brogaard2018}, namely KIC$5786154$, KIC$7037405$, KIC$8410637$, KIC$8430105$ and KIC$9970396$. It should be noted that in the present study, we concern ourselves with the stability of the mean density inversions. A full assessment of the accuracy of seismic masses and radii using individual frequencies would require a better assessment of the surface effect corrections, the importance of the helium abundance and of the atmospheric models used for the whole sample of eclipsing binaries available, which is beyond the scope of this study. 

\subsection{Peak-bagging and determination of individual frequencies}
Mode frequencies were estimated by performing bespoke mode fitting to each of the stars. Using the full set of available {\it Kepler} photometric data, we computed the estimate of the frequency power spectrum following \cite{2011MNRAS.414L...6G}. We located the modes of oscillation using visual inspection and determined the mode identification (i.e., selected the radial and quadrupole modes). We also checked the consistency of the detection with existing red giant measurements following \cite{2016AN....337..774D}. In order to determine the frequencies for the radial modes, we fitted each pair of radial and quadrupole modes with a sum of Lorentzians.  For full details of the mode fitting method, we refer to \cite{2016MNRAS.456.2183D}. Moreover, because of the binary nature of the targets, we have not applied a frequency correction to the mode frequencies normally applied to remove the Doppler shift as a result of the line-of-sight velocity \cite{2014MNRAS.445L..94D}.

\subsection{Forward modelling and inversion results}

The forward modelling process was carried out with the AIMS software, using the same grid as in the numerical exercises on artificial targets. Three separate runs were made to check for convergence and reliability of the process. The modelling was performed using only $\left[\mathrm{Fe}/\mathrm{H}\right]$ and the individual frequencies using surface corrections from \citet{Ball1}. For KIC$8430105$ and KIC$8410637$, a two terms surface correction was considered, while a single term correction was considered for the other targets whose spectrum contained much less oscillation modes. The masses and radii from this forward modelling process are given in Table \ref{tabMRForward} alongside the values from \citet{Gaulme2016} and \citet{Brogaard2018}. Again, we mention that they only consist in a preliminary modelling result, as other constraints such as effective temperature, luminosity and/or radius could change these values to a certain extent and that a more thorough study using various physical ingredients is required for a full assessment of the robustness of these masses and radii determinations. It should thus be noted that the reference models considered here used Eddington grey atmospheres and hence do not reproduce the effective temperature of the RGB. 

\begin{table*}
\caption{Masses and radii for the \textit{Kepler} eclipsing binaries from seismic forward modelling and from eclipses measurements ($\mathrm{EB-B}$ denotes values from \citet{Brogaard2018} and $\mathrm{EB-G}$ from \citet{Gaulme2016}).}
\label{tabMRForward}
  \centering
\begin{tabular}{r | c | c | c | c | c }
\hline
&KIC$7037405$ & KIC$5786154$& KIC$8430105$& KIC$8410637$& KIC$9970396$\\
\hline 
M$_{\mathrm{Sis}}$ (M$_{\odot}$) & $1.29\pm0.04$ & $1.12 \pm 0.05$ & $1.23\pm0.06$ & $1.53\pm0.06$& $1.15\pm 0.02$\\ 
R$_{\mathrm{Sis}}$ (R$_{\odot}$) & $14.19\pm0.17$ & $11.54 \pm 0.16$ & $7.45\pm0.11$ & $10.70\pm0.15$& $7.91\pm 0.06$ \\ 
M$_{\mathrm{EB-B}}$ (M$_{\odot}$) & $1.17 \pm 0.02$& $/$ & $/$ & $/$&$1.178 \pm 0.015$\\ 
R$_{\mathrm{EB-B}}$ (R$_{\odot}$) & $14.000 \pm 0.093$ & $/$ & $/$ & $/$&$8.035 \pm 0.074$\\ 
M$_{\mathrm{EB-G}}$ (M$_{\odot}$) & $1.25 \pm 0.03$ & $1.06 \pm 0.06$ & $1.31 \pm 0.02$& $1.56 \pm 0.03$& $1.14 \pm 0.02$ \\ 
R$_{\mathrm{EB-G}}$ (R$_{\odot}$) & $14.1\pm0.2$ & $11.4 \pm 0.2$ & $7.65\pm0.05$  & $10.7\pm0.1$& $8.0\pm 0.2$ \\
\hline
\end{tabular}
\end{table*}

The inversion results for the mean density of each target are given in Table \ref{tabINVEEB}, where we considered three cases for each reference model. First, we carried out the inversion without applying any surface correction to the frequencies, second, we applied the  \citet{Ball1} correction as determined by the forward modelling in AIMS, third, we applied the empirical \citet{Sonoi} correction on the frequencies. All these corrections were applied before carrying out the inversion, as including them directly in the cost function of the SOLA method does not allow for an accurate reproduction of the target function of the inversion and leads to strong biases.

From Table \ref{tabINVEEB}, we can see that the surface effect corrections can induce variations of up to $1$ per cent of the mean density of the star. This result confirms the fact that the errors of seismic inversions cannot be reliably determined only by the propagation of the observed uncertainties, as other effects such as model-dependency and effects of surface corrections will dominate the uncertainties. Similar tests were performed using a few models around the best reference determined by AIMS and the results remained within this $1$ per cent interval, defining the total variation of the inversion results. Therefore, taking into account potential further model-dependencies and uncertainties, we can consider that in these test cases, the mean density has been determined with a precision of $\pm 1.5$ per cent, considering a conservative interval $3$ times larger than that determined by the mean density inversion. This also implies that within these very conservative uncertainties, all reference models agreed with the inversion results.
 
\begin{figure}
	\centering
		\includegraphics[width=7cm]{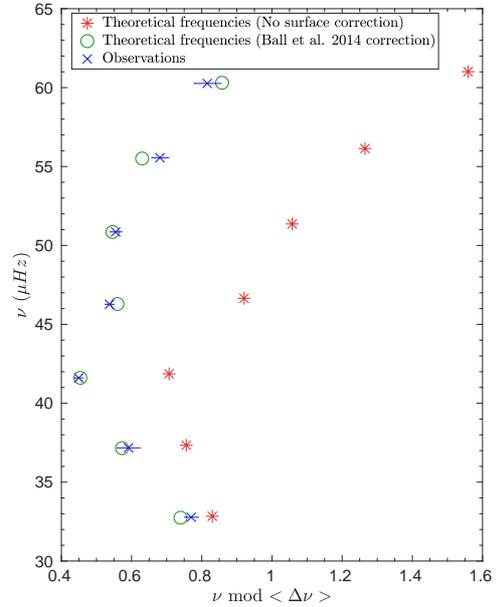}
	\caption{Echelle diagram of KIC$8410637$ illustrating the observed, theoretical frequencies (both corrected for surface effects using \citet{Ball1} formula and uncorrected).}
		\label{FigEchelle}
\end{figure}

This is no surprise, as most of the frequencies were very well fitted by the forward modelling process, as can be illustrated by the echelle diagram computed for KIC$8410637$, plotted in Figure \ref{FigEchelle} and showing the good agreement between theoretical and observed frequencies once surface effects corrections are included.

\begin{table*}
\caption{Inverted mean densities for the \textit{Kepler} eclipsing binaries of this study.}
\label{tabINVEEB}
  \centering
\begin{tabular}{r | c | c | c | c }
\hline
& $\bar{\rho}_{\mathrm{Ref}}$ $10^{-3}$(g/cm$^{3}$)& $\bar{\rho}^{\mathrm{No Surf}}_{\mathrm{Inv}}$ $10^{-3}$(g/cm$^{3}$)& $\bar{\rho}^{\mathrm{Ball}}_{\mathrm{Inv}}$ $10^{-3}$(g/cm$^{3}$)& $\bar{\rho}^{\mathrm{Sonoi}}_{\mathrm{Inv}}$ $10^{-3}$(g/cm$^{3}$)\\
\hline 
KIC$8430105$ & $4.209$ & $4.166 \pm 2\times 10^{-3}$ & $4.205 \pm 2 \times 10^{-3}$ & $4.199 \pm 2 \times 10^{-3}$ \\ 
KIC$5786154$ & $1.0275$ & $1.0254 \pm 5 \times 10^{-4}$ & $1.0309 \pm 5 \times 10^{-4}$ & $1.0342 \pm 5 \times 10^{-4}$ \\ 
KIC$7037405$ & $0.6374$& $0.6432 \pm 6 \times 10^{-4}$ & $0.6464 \pm 6\times 10^{-4}$ & $0.6482\pm 6 \times 10^{-4}$\\ 
KIC$8410637$ & $1.768$ & $1.762\pm 1\times 10^{-3}$ & $1.785 \pm 1 \times 10^{-3}$ & $1.770\pm 1 \times 10^{-3}$\\ 
KIC$9970396$ & $3.287$ & $3.317\pm 1 \times 10^{-3}$ & $3.329 \pm 1 \times 10^{-3}$ & $3.333 \pm 1 \times 10^{-3}$ \\ 
\hline
\end{tabular}
\end{table*}

\begin{figure*}
	\centering
		\includegraphics[width=17.8cm]{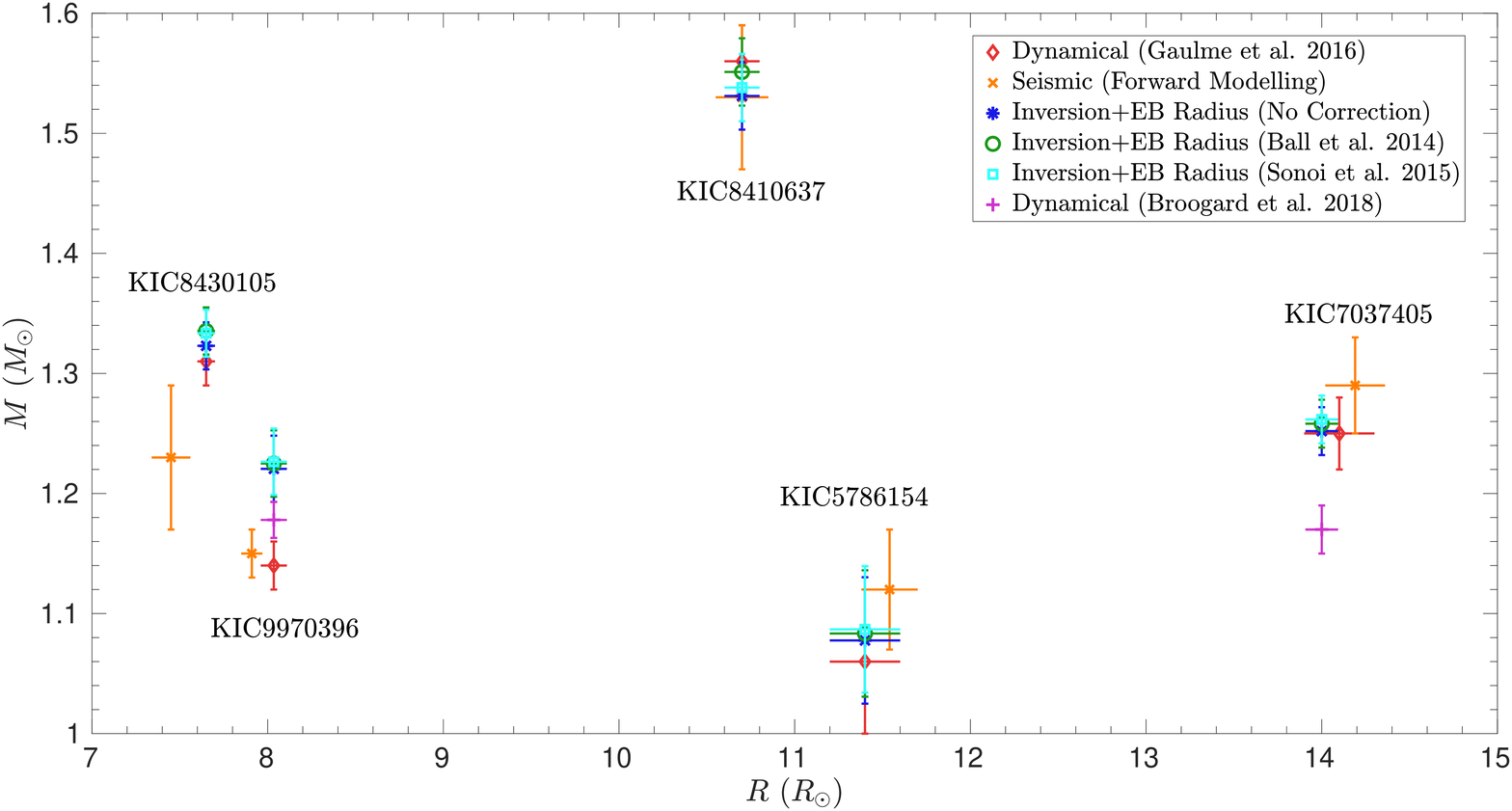}
	\caption{Masses and radii values obtained from eclipses observations, seismology, and combining the inverted mean densities using various surface corrections to the dynamical radii measures.}
		\label{FigMR}
\end{figure*}

As such, this implies that seismic inversions of the mean density can in these cases play a first role of confirmation of the quality of the fit, but can also be included in a second step of forward modelling to determine better masses for red giants. Indeed, using individual frequencies directly as constraints can lead to cost functions dominated by seismic information and unrealistically precise determinations of fundamental parameters of stars. Hence, using the inverted mean density directly as a kind of observational constraint, as done for example for the $16$Cyg binary system in \citet{BuldgenCygA,BuldgenCygB}, alongside luminosities or radii determined by Gaia may lead to more accurate  masses of red giants and clump stars. To illustrate this, we show in Figure \ref{FigMR} the mass values obtained from the inverted mean densities and radii values from the eclipses observations. We selected the most precise determinations of radii between \citet{Brogaard2018} and \citet{Gaulme2016}. This is not a major concern since the radii values agree very well with each other for the targets present in both studies. We note that for the case of KIC$7037405$ and KIC$9970396$, the mean density inversions lead to an unambiguous correction which leads to a very good agreement with the dynamical mass of \citet{Gaulme2016} for KIC$7037405$. Further investigations show that the mean density derived from the dynamical values of the mass and radius from \citet{Brogaard2018} differs by around $4 \pm 2.5$ per cent with the ones determined in this study. Similarly, the values determined from global seismic indices using the PARAM software \citep{Rodrigues2017} also show discrepancies of around $2$ per cent. For this particular target, the closest value to the inverted one is found using the mass and radii values from the corrected scaling relations in Table $4$ of \citet{Brogaard2018}, while the uncorrected scaling relations give a disagreement of $4$ per cent. 

As for KIC$9970396$, a slight disagreement remains and further modelling using tighter constraints is required to determine whether the slightly higher mass value found by seismology is due to model-dependencies or if the origin is to be found elsewhere. In this case, the mean density value determined from the eclipses is in agreement within $1 \sigma$ (for this case around $3$ per cent) with the value determined here. The value determined from the corrected scaling relations shows an already quite good agreement. However, the mean density value determined from the uncorrected scaling relations shows a larger disagreement of approximately $6$ per cent whereas the values determined from PARAM differ by approximately $3$ per cent with the inverted results and significantly from the mean density value from the eclipses. This seems to indicate that overall, using the entire information of the frequency spectrum leads to the best agreement with the dynamical values. 

We can see in this Figure that for all cases, using the inverted mean density and the dynamical radii further improve the results. However, most of the improvement does not stem from the improvement of the mean density as the corrections remained quite small, but rather from the use of the radii values from the eclipses. This implies that the use of classical constraints such as precise luminosities or radii will lead a major role in the determination of reliable masses for red giants and red clump stars. 

\section{Conclusion}\label{sec:Conclu}

In this study, we have demonstrated the feasibility and robustness of seismic inversions of the mean density of red giant and red clump stars using only a few observed radial modes, providing an extension to the framework of the approach initially applied to main-sequence solar-like stars \citep{Reese, Buldgen}. We have started by introducing the approach to the inverse problem in Sect. \ref{sec:inversion} and applied it in extensive numerical tests using calibration techniques based on effective temperature and luminosity in Sect. \ref{sec:TeffL} as well as seismic constraints \ref{sec:SismoCons}. We have analysed the possibility of carrying out inversions for the mean density both for red giant branch and clump stars, as well as trying to determine the mean density of a misidentified clump star. In the last case, the inversion proved to be inaccurate and thus the method provided here is only valid for unambiguously identified red giant branch and clump stars. This last point also proves the need for a reliable reference model before carrying out the inversion. This weakness has also been observed for other numerical exercises were the so-called cross-term errors could contribute very significantly to the total error budget of the inversion and even dominate other errors. This is radically different from inversions on the main-sequence and is due to both the small number of modes and the large radial extent of the ionization zones, leading to more widespread differences in $\Gamma_{1}$ from one model to the other. 

In addition to testing the use of seismic constraints, we also carried out in Sect. \ref{sec:SurfEff} inversions for an artificial target from \citet{Sonoi} including an atmospheric model from an averaged hydrodynamical simulation. For this target, we used both adiabatic and non-adiabatic frequencies to carry out the inversion. These tests illustrated the impact of surface effects and the importance of correcting them. It should be noted that none of the current methods seemed to work perfectly, as error compensations were sometimes observed and that directly implementing the corrections as free parameters in the SOLA method provided inaccurate fits of the target function of the inversion.

In Sect. \ref{sec:ObsBin}, we carried out mean density inversions on observed red giants in eclipsing binary systems from the studies of \citet{Gaulme2016} and \citet{Brogaard2018}. We first showed that using individual frequencies, corrected for surface effects using the approach of \citet{Ball1}, lead to a much better agreement in terms of mass and radius than simply using scaling relations or global seismic indices. Furthermore, we showed that the mean density inversions could provide either further small corrections to the mean density obtained from forward modelling or an additional verification step. Combining the inverted mean density to the radii determinations from eclipses provided an overall good agreement in terms of masses for these stars. However, these test cases also demonstrate the importance of classical constraints for the modelling of red giants, since the most significant improvement came from using the dynamical radii values. Indeed, the mean density of these stars was already very accurately determined through forward modelling. 

A couple of conclusions can be drawn from this last point. Firstly, that pure seismic modelling using only radial modes might lead to degeneracies since the frequencies will be mostly sensitive to the mean density of the star. Secondly, directly using the individual frequencies might lead to underestimated uncertainties and the seismic constraints might dominate the classical constraints. Therefore, using directly an inverted value for the mean density, alongside a precise and accurate value for the luminosity, the $\left[\mathrm{Fe/H}\right]$ and the radius might provide a more direct and balanced approach to the modelling of red giant stars. Other seismic constraints could also be used, such as the asymptotic period spacing or ratios of radial oscillation frequencies as is done for Cepheids and other classical pulsators. The application of such approaches to an extended set of eclipsing binaries will provide a unique opportunity to test the reliability of seismic modelling and the importance of classical constraints. 

In the near future, using such approaches alongside constraints from the second Gaia data release will help better understand the properties of red giants. Providing more accurate masses is indeed crucial to determine the properties of various stellar populations in the Galaxy but also for example to pinpoint the properties of additional mixing at the base of the convective envelope, manifesting itself through the so-called RGB bump \citep{Alongi91,Cassisi2011,Khan2018}. In stellar clusters, accurate masses could also be used to characterize mass loss on the red giant branch \citep{HandbergRG}, one of the major issues in current stellar evolution models.

\section*{Acknowledgements}

AM and GB acknowledge support from the ERC Consolidator Grant funding scheme ({\em project ASTEROCHRONOMETRY}, G.A. n. 772293). We gratefully acknowledge the support of the UK Science and Technology Facilities Council (STFC). S.J.A.J.S. is funded by ARC grant for Concerted Research Actions, financed by the Wallonia-Brussels Federation. TS acknowledges support from JSPS KAKENHI Grant Number 17J00631. This article made use of an adapted version of InversionKit, a software developed in the context of the HELAS and SPACEINN networks, funded by the European Commissions's Sixth and Seventh Framework Programmes. 




\bibliographystyle{mnras}
\bibliography{biblioarticleGiants} 

\bsp	
\label{lastpage}
\end{document}